\documentclass[aps,prb,twocolumn,showpacs,amsmath,amssymb]{revtex4}
\usepackage[latin2]{inputenc}
\usepackage{amsmath}
\usepackage{graphicx}
\usepackage{multirow}
\usepackage{dcolumn}
\usepackage{epstopdf}

\newcommand{\ba}{\begin{eqnarray*}}
\newcommand{\ea}{\end{eqnarray*}}
\newcommand{\baa}{\begin{eqnarray}}
\newcommand{\eaa}{\end{eqnarray}}
\def\bar{\begin{array}}
\def\ear{\end{array}}
\def\LB{\left(}
\def\RB{\right)}

\def\u{\uparrow}
\def\d{\downarrow}
\def\pa{\uparrow\!\!\downarrow}
\def\s{\sigma}
\def\f{\frac}

\def\nn{\nonumber}
\def\R{\mathbf{R}}
\def\r{\mathbf{r}}
\def\U{\mathbf{U}}

\begin{document}

\title{Molecular geometric phase from the exact electron-nuclear factorization} 

\author{Ryan Requist}
\email{rrequist@mpi-halle.mpg.de}
\affiliation{
Max Planck Institute of Microstructure Physics, Weinberg 2, 06114 Halle, Germany 
}
\author{Falk Tandetzky}
\affiliation{
Max Planck Institute of Microstructure Physics, Weinberg 2, 06114 Halle, Germany
}
\author{E. K. U. Gross}
\affiliation{
Max Planck Institute of Microstructure Physics, Weinberg 2, 06114 Halle, Germany
}

\date{\today}

\begin{abstract}
The Born-Oppenheimer electronic wavefunction $\Phi_R^{BO}(r)$ picks up a topological phase factor $\pm 1$, a special case of Berry phase, when it is transported around a conical intersection of two adiabatic potential energy surfaces in $R$-space.  We show that this topological quantity reverts to a geometric quantity $e^{i\gamma}$ if the geometric phase $\gamma = \oint \mathrm{Im} \langle \Phi_R |\nabla_{\mu} \Phi_R\rangle \cdot d\R_{\mu}$ is evaluated with the conditional electronic wavefunction $\Phi_R(r)$ from the exact electron-nuclear factorization $\Phi_R(r)\chi(R)$ instead of the adiabatic function $\Phi_R^{BO}(r)$. A model of a pseudorotating molecule, also applicable to dynamical Jahn-Teller ions in bulk crystals, provides the first examples of induced vector potentials and molecular geometric phase from the exact factorization.  The induced vector potential gives a contribution to the circulating nuclear current which cannot be removed by a gauge transformation.  The exact potential energy surface is calculated and found to contain a term depending on the Fubini-Study metric for the conditional electronic wavefunction.
\end{abstract}

\pacs{03.65.Vf, % Phases: geometric, topol
31.30.-i % Born-Oppenheimer approximation
}

\maketitle

\section{Introduction}

The Born-Oppenheimer approximation underlies most calculations in condensed matter physics and chemistry.  Examples include thermal conductivity, lattice-mediated relaxation of excited electrons and optical properties of materials, as well as molecular scattering and rovibronic spectroscopy.  Since nuclei are much heavier than electrons, one can get a good approximation to the electron-nuclear wavefunction $\Psi(r,R) \approx \Phi_R^{BO}(r) \chi^{BO}(R)$ by assuming the nuclei are frozen and solving an electronic Schr\"odinger equation with the $R$-dependent Hamiltonian $\hat{H}^{BO} = \hat{T}_e +  \hat{V}_{ee} + \hat{V}_{en} + \hat{V}_{nn}$, which is the full electron-nuclear Hamiltonian with the nuclear kinetic energy removed.  The eigenvalue defines an adiabatic potential energy surface     
\begin{equation}
\mathcal{E}^{BO}(R) = \langle \Phi_R^{BO} | \hat{T}_{e} + \hat{V}_{ee} + \hat{V}_{en} + \hat{V}_{nn} |  \Phi_R^{BO} \rangle {,}
\end{equation} 
which is then used in the nuclear Schr\"odinger equation 
\begin{equation}
\sum_{\mu=1}^{N} \Big[ -\f{\hbar^2 \nabla_{\mu}^2}{2M_{\mu}} +  \mathcal{E}^{BO}(R) \Big] \chi^{BO}(R) = E \chi^{BO}(R) {.} \label{eq:BO:nucl}
\end{equation}
The adiabatic potential energy surface is an extremely useful concept which implicitly encapsulates all electronic terms (kinetic $T_{e}$, interaction $V_{ee}$ and electron-nuclear coupling $V_{en}$) and the nuclear interaction $V_{nn}$ in a single scalar function $\mathcal{E}^{BO}(R)$ under the assumption that the electronic wavefunction $\Phi_R^{BO}(r)$ stays in the ground state of the electronic Hamiltonian for all values of the nuclear coordinates $R$; we use the notations $R=(\R_1,\R_2,\ldots)$ and $r=(\r_1,\r_2,\ldots)$.  

A curious feature of the Born-Oppenheimer approximation is the occurrence of conical intersections between the potential energy surfaces of two or more electronic eigenstates in some polyatomic molecules.\cite{domcke2004,bersuker2006}  
The factor $\chi^{BO}(R)$ is then multivalued due to the non\-analyticity of the potential $\mathcal{E}^{BO}(R)$ at the point of intersection.  Since the full wavefunction $\Phi_R^{BO}(r) \chi^{BO}(R)$ must be a single-valued function of $R$, multivaluedness of $\chi^{BO}(R)$ implies that $\Phi_R^{BO}(r)$ is also multivalued, so it does not return to its original value if transported along a closed path in $R$-space encircling a conical intersection, but instead changes sign.
This sign change is due to the Longuet-Higgins phase.\cite{herzberg1963}  It is a special case of the Berry phase \cite{berry1984} because it only takes the values 0 and $\pi$.

Multiplication by a Dirac phase factor $\mathrm{exp} \f{i}{\hbar}\int \mathbf{A}_{\mu}\cdot d\R_{\mu}$ which compensates the sign change makes $\Phi_R(r)$ single-valued.  If this choice of phase is made, $\mathrm{Im} \langle \Phi_R |\nabla_{\mu} \Phi_R\rangle$ is no longer zero, and Eq.~(\ref{eq:BO:nucl}) must be replaced by\cite{mead1979,moody1986}
\begin{equation}
\sum_{\mu=1}^{N} \Big[ \f{(-i\hbar \nabla_{\mu} + \mathbf{A}_{\mu}^{BO})^2}{2M_{\mu}} +  \mathcal{E}^{BO}(R) \Big] \chi^{BO}(R) = E \chi^{BO}(R) {,}\label{eq:BO:mead}
\end{equation}
where $\mathbf{A}_{\mu}^{BO} = \hbar\: \mathrm{Im} \langle \Phi_R^{BO} |\nabla_{\mu} \Phi_R^{BO}\rangle$ is the \textit{induced} vector potential introduced by Mead and Truhlar.\cite{mead1979}

Nonadiabatic coupling between electronic eigenstates causes corrections to the Born-Oppenheimer approximation, which are usually included through the expansion\cite{born1954,baer2002,baer2006}
\begin{align}
\Psi(r,R) = \sum_n \Phi_{n}(r,R) \chi_n(R) {,} \label{eq:adiabatic:expansion}
\end{align}
i.e.~through a sum of Born-Oppenheimer-like factors, one for each stationary state $\Phi_{n}(r,R)$.  Surprisingly, it is not actually necessary to depart from the Born-Oppenheimer single-product form to include non\-adiabatic effects.  In fact, the exact wavefunction can be factored into a single product $\Phi_R(r) \chi(R)$ called the \textit{exact electron-nuclear factorization}. \cite{hunter1975,gidopoulos2014,abedi2010,abedi2012}  Like the Born-Oppenheimer ansatz, the exact factorization can be used to define scalar and vector  potentials $\mathcal{E}(R)$ and $\mathbf{A}_{\mu}(R)$.

Using these exact potentials in place of $(\mathcal{E}^{BO},\mathbf{A}_{\mu}^{BO})$ in Eq.~(\ref{eq:BO:mead}) yields the exact nuclear density and nuclear current density of the state $\Psi(r,R)$.\cite{abedi2010}  Since it integrates all nonadiabatic electronic effects into a single potential energy surface and vector potential, the exact factorization provides an intuitive and economical description of quantum nuclear dynamics.  From this standpoint, the following two questions are relevant.  How does the exact potential energy surface differ from the adiabatic ones, e.g.~do conical intersections persist?  Do the vector potential and Longuet-Higgins phase connected with the nonanalyticity at the point of conical intersection remain nonzero in the exact factorization?  If the vector potential can be made to vanish by a gauge transformation, then the potential energy surface is all one needs to describe the nuclear motion.  If instead the molecular geometric phase is nontrivial, then the nuclear Schr\"odinger equation must contain induced vector potentials.  Recent work found a case in which the Longuet-Higgins phase of $\pi$ becomes zero in the exact factorization.\cite{min2014} However, the model studied does not have the degeneracy of the classic Jahn-Teller models\cite{jahn1937} of pseudorotating molecules\cite{jahn1937,moffitt1957,longuet-higgins1958,child1961,herzberg1963,mead1979} and transition metal ions in bulk crystals\cite{vanvleck1939,abragam1950,child1961,slonczewski1963,obrien1964,englman1972,ham1972,ham1987} and therefore leaves room for further investigation.

Within the Born-Oppenheimer approximation, Berry phase effects have been extensively studied in pseudorotating molecules such as Na$_3$\cite{mead1979,delacretaz1986,whetten1986,zwanziger1987,coccini1988,mead1992,koizumi1994,kendrick1997,vonbusch1998,allen2005,ryabinkin2013,lee2013},  hydrogen exchange reactions\cite{mead1980a,wu1991,kendrick2015} and fullerene ions and crystals.\cite{varma1991,ihm1994,auerbach1994,manini1994,hands2006,iwahara2013}  
Berry phase effects are also relevant to some dynamical Jahn-Teller systems which have been investigated recently.\cite{perebeinos2005,bersuker2007,haupricht2010,sanz-ortiz2010,popovic2012,atanasov2012,kovaleva2013,pae2014} In many of these cases, the exact factorization could provide an interesting alternative to conventional approaches based on the adiabatic expansion of Eq.~(\ref{eq:adiabatic:expansion}).  Because the nuclear wavefunction is determined by a single potential energy surface and vector potential, rather than an infinite set of coupled adiabatic potential energy surfaces, the exact factorization scheme might prove to more efficient than traditional approaches to coupled electron-nuclear dynamics if accurate approximations can be found for $\mathcal{E}$ and $\mathbf{A}_{\mu}$.

A two-mode vibronic model has been studied in the context of the exact factorization to explore features of the exact potential energy surface, such as spikes that occur near nodes of the adiabatic nuclear eigenfunctions.\cite{chiang2014}  Such spikes were previously observed in the 1D potential energy surfaces of diatomic molecules.\cite{czub1978,hunter1980,hunter1981,cassam-chenai2006,lefebvre2015}  Here, we focus on the exact vector potential in a model pseudorotating triatomic molecule known to have a degenerate ground state and nonzero Longuet-Higgins phase in the adiabatic approximation.\cite{allen2005}  The presence of degeneracy, which is maintained in the exact approach, is pivotal: the exact factorization can be applied to any state in the ground state manifold, and the potential energy surface and vector potential will depend on which state is chosen.  Any choice breaks the unitary symmetry of the ground state manifold and leads to a corresponding symmetry breaking of the exact potential energy surface and/or vector potential.  Our main result is that the discrete topological Longuet-Higgins phase, 0 or $\pi$, becomes a full geometric phase $e^{i\gamma}$ in the exact approach.  

Section~\ref{sec:model} introduces our model pseudorotating triatomic molecule.  After solving the full electron-nuclear Schr\"odinger equation in Sec.~\ref{sec:solution}, we evaluate the vector potential and molecular geometric phase in Sec.~\ref{sec:geometricphase}, and the exact potential energy surface in Sec.~\ref{sec:PES}.  The Berry curvature and a Riemannian metric derived from the conditional electronic wavefunction are unified in a quantum geometric tensor in Sec.~\ref{sec:QGT}.  

\section{Model pseudorotating molecule \label{sec:model}}

Our model pseudorotating molecule consists of three hydrogen-like atoms, whose nuclei are assumed to be distinguishable to avoid the complication of nuclear exchange symmetry.  This models a molecule like LiNaK, but we further assume the nuclear masses are equal.  We will follow, as closely as possible, the notations of Ref.~\onlinecite{allen2005}, where the model was introduced.  The well known $E\otimes e$ Jahn-Teller system can be obtained by the truncation to two electronic levels, which would be a good approximation when the electronic level spacing is much larger than the characteristic vibrational energy.

The electronic degrees of freedom are described within the truncated Hilbert space formed by three valence electrons occupying three $s$-like orbitals, one per atom, while the full real space $R$-dependence of the nuclear states is retained.  The Hamiltonian is
\begin{equation}
\hat{H} = \hat{T}_n + \hat{V}_{nn} + \hat{H}_{en}, \label{eq:H}
\end{equation}
where $T_n$ is the nuclear kinetic energy, $V_{nn}$ is the nuclear interaction energy and 
\begin{equation}
\hat{H}_{en} = -\sum_{n\s} \LB t_{n,n+1}(R) c_{n\s}^{\dag} c_{n+1\s} + H.c. \RB \label{eq:Hen:1}
\end{equation}
subsumes electronic kinetic energy and electron-nuclear coupling.  The electron-nuclear coupling is represented in the $R$-dependence of the hopping amplitudes $t_{n,n+1}$.  Electron-electron interactions are neglected; their effect has been studied in the adiabatic limit.\cite{allen2005}  
\begin{figure}[tb!]
\includegraphics[width=0.6\columnwidth]{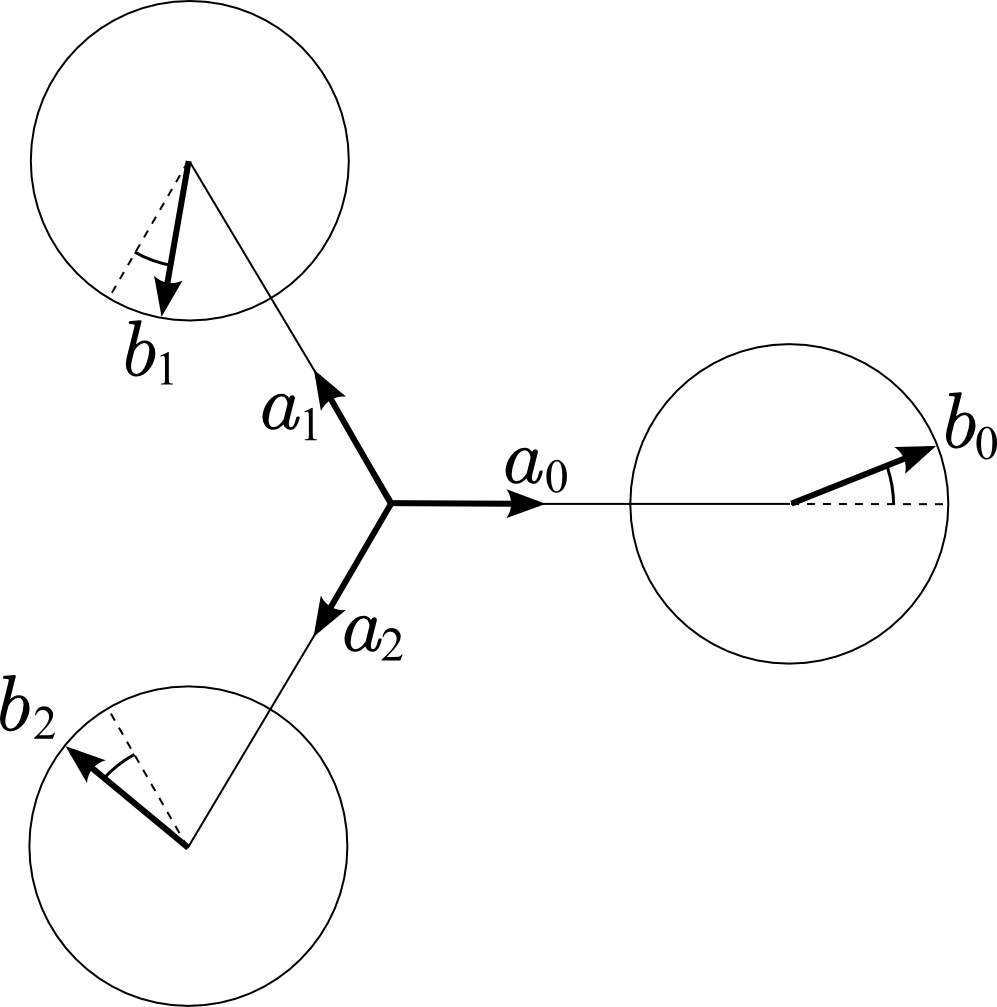}
\caption{Two sets of unit vectors $\mathbf{a}_n$ and $\mathbf{b}_n$ used to specify the geometry of the distorted triatomic molecule.}
\label{fig:triangle}
\end{figure}

The key simplifying assumption of the model is the truncation of the Taylor expansions of $H_{en}$ and $V_{nn}$ with respect to distortions of the equilateral triangle geometry (see Fig.~\ref{fig:triangle}).  For the hopping amplitude $ t_{n,n+1}(R)$ in Eq.~(\ref{eq:Hen:1}), we keep only the linear term
\begin{align}
t_{n,n+1}(R) &= t_0 - \f{g}{\sqrt{3}} \big( |\R_{n+1}-\R_n| - \sqrt{3}R_0 \big) \nonumber \\
&= t_0 - \f{g}{3} ( \U_{n+1}-\U_n ) \cdot ( \mathbf{a}_{n+1}-\mathbf{a}_n ) {,}
\label{eq:linearization}
\end{align}
where $g$ is the electron-nuclear coupling constant, $\sqrt{3} R_0$ is the internuclear separation in the equilateral configuration, $\U_n$ is the deviation of nuclear coordinate $\R_n$ from its equilateral position, i.e.~$\mathbf{U}_n = \R_n-R_0 \mathbf{a}_n$, and $\mathbf{a}_n$ is the set of unit vectors shown in Fig.~\ref{fig:triangle}.  We choose $t_0=1$ to be our characteristic unit of energy.  The term $V_{nn}$ models internuclear repulsion and some part of the electron-nuclear attraction.  We assume harmonic spring interactions by expanding to second order in $\mathbf{U}_n$:
\begin{align}
\hat{V}_{nn} &= \f{K}{2} \sum_{n=1}^3 \big( | \R_{n+1} - \R_n | - \sqrt{3} R_0 \big)^2 \nn \\
&= \f{K_1}{2} Q_1^2 + \f{\mathcal{K}}{2} Q^2 {,}
\end{align}
where $Q_1$ and $Q$ are internal nuclear coordinates defined below, and we have introduced the effective spring constants $K_1=3K$ and $\mathcal{K} = \f{9}{2} K$.

The three nuclear coordinates are determined by three center of mass coordinates, three Euler angles and three internal coordinates.   The nuclear coordinates in the laboratory frame are
\begin{align}
\R_n^{\rm lab} = \mathbf{R}_{cm} + \mathcal{T} \,\R_n^{\rm mol}{,} 
\end{align}
where $\mathbf{R}_{cm}$ is the nuclear center of mass, $\mathcal{T}$ is an SO(3) rotation and $\R_n^{\rm mol}$ are molecular frame coordinates.  The molecular frame coordinates can be parameterized by three normal mode amplitudes $(Q_1,Q_2,Q_3)$; $Q_1$ is the breathing mode; $Q_2$ and $Q_3$ are the conventional symmetric and asymmetric bending modes.\cite{vanvleck1939,mead1992}  In polar coordinates $Q=\sqrt{Q_2^2+Q_3^2}$, $\eta=\tan^{-1}(Q_3/Q_2)$, we have
\begin{align}
\R_n^{\rm mol} &= (R_0+Q_1) \mathbf{a}_n + Q \mathbf{b}_n \nn \\ \mathbf{a}_n &= \Big( \cos\Big(\f{2\pi n}{3}\Big), \sin\Big(\f{2\pi n}{3}\Big), 0 \Big) \nn \\
\mathbf{b}_n &= \Big( \cos\Big(\eta - \f{2\pi n}{3}\Big), \sin\Big(\eta - \f{2\pi n}{3}\Big), 0 \Big) {.}
\label{eq:R:mol}
\end{align}
The $\eta$-dependent unit vectors $\mathbf{b}_n$ are shown in Fig.~\ref{fig:triangle}.  The nuclear kinetic energy operator is 
\begin{align}
\hat{T}_n^{\rm mol} = -\f{\hbar^2}{2\mathcal{M}} \Big( \f{1}{Q} \f{d}{dQ} \Big(Q \f{d}{dQ}\Big) + \f{1}{Q^2} \f{d^2}{d\eta^2} + \f{d^2}{dQ_1^2} \Big) {.} \label{eq:kinetic}
\end{align}
The total nuclear mass $\mathcal{M} = 3M$ appears here because $(Q_1,Q,\eta)$ describe collective nuclear motion.  As one further simplification, we neglect the last term in Eq.~(\ref{eq:kinetic}) and assume that $Q_1$ is frozen to its equilibrium value.  Since $Q_1$ is a fully symmetric mode, this simplification will have no effect on our qualitative conclusions concerning the molecular geometric phase.  The Hamiltonian is now fully defined, and we proceed to calculate its eigenstates.

\section{Exact solution of the model \label{sec:solution}}

The model will be solved by exact diagonalization after the Hamiltonian matrix elements are calculated in an electron-nuclear product basis. 

We start by defining a complete set of electronic states.  The truncated 3-electron Hilbert space is 20-dimensional, but since $\hat{\mathbf{S}}^2$ and $\hat{S}_z$ commute with the Hamiltonian, we focus on the 8-dimensional sector with spin quantum numbers $S=\f{1}{2}$ and $S_z=\f{1}{2}$.  
The basis states are constructed from the ket $|\rm core\rangle$ representing the inert core electrons by acting with the creation operators of single-particle orbitals $|\phi_k\rangle = \f{1}{\sqrt{3}} \sum_{n=0,1,2} e^{i 2\pi kn/3}|n\rangle$, $k=1,0,-1$, e.g.~$| \!\u\: \pa 0 \rangle = c_{+1\u}^{\dag} c_{0\u}^{\dag} c_{0\d}^{\dag} |\rm core\rangle$, where the position of the spin in the ket represents the orbital it occupies with the ordering convention $k=1,0,-1$.  Thus, we choose the following eight basis states:
\begin{align}
|a \rangle &= | \!\u\; \pa 0 \rangle \nn \\ 
|b \rangle &= | \,0 \pa\; \u\rangle \nn \\[0.2cm]
|c \rangle &= | \!\pa 0 \u \rangle  \nn \\ 
|d \rangle &= | \!\u\, 0 \pa \rangle \nn \\[0.2cm]
|e \rangle &= | \,0 \u\; \pa\rangle \nn \\
|f \rangle &= | \!\pa\; \u 0\rangle \nn \\[0.1cm]
|g \rangle &=  -\f{i}{\sqrt{6}} \Big( 2 |\u\; \d\; \u\rangle - |\u\; \u\; \d\rangle - |\d\; \u\; \u\rangle \Big) \nn \\
|h \rangle &= -\f{1}{\sqrt{2}} \Big( 0 |\u\; \d\; \u\rangle + |\u\; \u\; \d\rangle - |\d\; \u\; \u\rangle \Big) {.}
\label{eq:basis:orb}
\end{align}

The nuclear basis states in the polar coordinate representation are $\f{1}{\sqrt{2\pi}} \rho_{nm}(Q) e^{im\eta}$ with radial and azimuthal quantum numbers $n$ and $m$.  As a complete basis of radial functions, we choose the normalized radial eigenfunctions of the isotropic 2D harmonic oscillator
\begin{align*}
\rho_{nm}(Q) = \f{\sqrt{2\lambda N!}}{\sqrt{(N+|m|)!}} \big(\lambda Q^2 \big)^{|m|/2} e^{-\lambda Q^2/2} L_{N|m|}(\lambda Q^2),
\end{align*}
where $N=(n-|m|)/2$, $\lambda = \sqrt{\mathcal{K}\mathcal{M}/\hbar^2}$ and $L_{Nm}$ is the associated Laguerre polynomial, defined by the differential equation
\begin{align*}
x y^{\prime\prime} + (m+1-x) y^{\prime} + Ny = 0 {.}
\end{align*}
 
The next step is to evaluate the Hamiltonian matrix elements in the electron-nuclear basis $|anm\rangle=|a\rangle|nm\rangle$: 
\begin{align}
\langle a_1 n_1 m_1 | \hat{H} | a_2 n_2 m_2 \rangle &= \delta_{a_1 a_2} \langle n_1 m_1 | \hat{T}_n + \hat{V}_{nn} | n_2 m_2 \rangle \nn \\
&+ \langle a_1 n_1 m_1 | \hat{H}_{en} | a_2 n_2 m_2 \rangle {.} \label{eq:H:matrix}
\end{align}
All matrix elements can be evaluated analytically and are reported in App.~\ref{app:A}.

Solving the model by exact diagonalization confirms that the ground state is doubly degenerate for certain values of the parameters $M$, $K$ and $g$.  
Before going on to investigate the molecular geometric phase, it is instructive to examine the symmetries of the model. 

Threefold symmetry is responsible for several special properties of the model.  A combined symmetry operator $\hat{C}_3 = \hat{C}_{3e} \hat{C}_{3\eta}$, where $\hat{C}_{3e}$ is a threefold permutation of the electrons ($0\rightarrow 1\rightarrow 2\rightarrow 0$) and $\hat{C}_{3\eta}$ is a threefold \textit{pseudo}rotation of the nuclei ($\eta\rightarrow\eta-2\pi/3$), commutes with the model Hamiltonian in Eq.~(\ref{eq:H}).  This symmetry is what remains of the rotational symmetry of the original real-space Hamiltonian in our model in which overall molecular rotations are not included.  To see this, we note that if $\hat{C}_3$ is combined with a three-fold permutation of the nuclei $0\rightarrow 1\rightarrow 2\rightarrow 0$, denoted by $\hat{C}_{3n}$, then the combined operation $\hat{C}_{3}\hat{C}_{3n}$ corresponds to a $2\pi/3$ rotation of the entire molecule, which is subgroup of the full rotational symmetry of the original Hamiltonian.

The operator $\hat{C}_{3e}$ shifts the electrons forward by one site, e.g.~$\hat{C}_{3e}|\!\!\u\: \pa \!0 \rangle \nn = | \,0 \!\u\: \pa\rangle \nn$ for an arbitrary ket expressed in the $(0,1,2)$ site basis.  The basis functions in Eq.~(\ref{eq:basis:orb}) were chosen to be eigenstates of $\hat{C}_{3e}$:  $|a\rangle$, $|c\rangle$, $|e\rangle$ have eigenvalue $e^{-i2\pi/3}$; $|b\rangle$, $|d\rangle$, $|f\rangle$ have eigenvalue $e^{i2\pi/3}$; $|g\rangle$ and $|h\rangle$ have eigenvalue $1$.  Similarly, the operator $\hat{C}_{3\eta}$ rotates the nuclei by $-2\pi/3$ in $(Q_2,Q_3)$ space as shown in Eq.~(\ref{eq:C3n}) below.  

The threefold symmetry group generated by $\hat{C}_3$ has two irreducible representations -- a symmetric singlet $A$ and a doublet $E$.  Being a doublet, the ground state belongs to the $E$ representation.  We can choose two orthogonal states from the ground state manifold that transform into themselves up to a phase under $\hat{C}_3$.  These states have eigenvalues $e^{\mp i 2\pi/3}$ and will be labeled $|\Psi_{\pm}\rangle$ with $+/-$ indicating counterclockwise/clockwise nuclear current.  Only $|anm\rangle$ states with the same eigenvalue of $\hat{C}_3$ are coupled by the Hamiltonian.  Hence, the state $|\Psi_+\rangle$, for instance, has the following structure:
\begin{align}
|\Psi_+\rangle &= \Gamma_a |a\rangle + e^{i\eta} \Gamma_b |b\rangle + \Gamma_c |c\rangle + e^{i\eta} \Gamma_d |d\rangle \nn \\ 
&+ \Gamma_e |e\rangle + e^{i\eta} \Gamma_f |f\rangle 
+ e^{-i\eta} \big( \Gamma_g |g\rangle + \Gamma_h |h\rangle \big) {,}
\label{eq:form:exact}
\end{align}
where $\Gamma_{\alpha}=\Gamma_{\alpha}(Q,\eta)$ are periodic functions of $\eta$ with period $2\pi/3$, i.e.
\begin{equation}
\Gamma_{\alpha}(Q,\eta) = \sum_{m=-\infty}^{\infty} \Gamma_{\alpha,3m}(Q) e^{i3m\eta} {.}
\end{equation}
The state $|\Psi_-\rangle$ is the complex conjugate of $|\Psi_+\rangle$.  The operator $\hat{C}_{3\eta}$ acts on the nuclear functions as
\begin{equation}
\hat{C}_{3\eta} \Gamma_{\alpha}(Q,\eta) = \Gamma_{\alpha}(Q,C_{3\eta}^{-1}\eta) = \Gamma_{\alpha}(Q,\eta + 2\pi/3) {.}
\label{eq:C3n}
\end{equation}
Thus, we immediately verify that $\hat{C}_3|\Psi_+\rangle = e^{-i2\pi/3} |\Psi_+\rangle$.   

To see how double-valued electronic and nuclear wavefunctions, $|\Phi_R^{BO}\rangle$ and $\chi(R)$, can emerge from the single-valued function in Eq.~(\ref{eq:form:exact}), consider the simultaneous $M\rightarrow\infty$ and $g\rightarrow 0$ limit with the condition $\hbar \Omega/\Delta=const$, where $\Delta=g^2/2\mathcal{K}$ is the Jahn-Teller stabilization energy.  In this limit, it can be shown that
\begin{align}
|\Psi_+\rangle &\rightarrow \f{ \rho(Q)}{\sqrt{2}} |a\rangle + \f{\rho(Q)}{\sqrt{2}} e^{i\eta}  |b\rangle \nn \\
&=  \rho(Q)  e^{i\eta/2} \Big( \f{1}{\sqrt{2}}e^{-i\eta/2} |a\rangle +\f{1}{\sqrt{2}} e^{i\eta/2} |b\rangle \Big) {.}
\label{eq:limit}
\end{align}
The function in parentheses is the real-valued electronic Born-Oppenheimer function $|\Phi_R^{BO}\rangle$, showing the characteristic sign change when $\eta$ increases by $2\pi$.  The nuclear factor is $\chi(Q,\eta) = \rho(Q) e^{i\eta/2}$,
where $\rho(Q)$ is approximately equal to a harmonic oscillator wavefunction centered at the minimum of the potential $\frac{\mathcal{K}}{2} Q^2 - gQ$.  The lowest order corrections to Eq.~(\ref{eq:limit}) in the $M\rightarrow\infty$ limit will be investigated in further detail elsewhere.  

Instead of $|\Psi_+\rangle$ and $|\Psi_-\rangle$, one can choose two real-valued orthogonal states from the ground state manifold, e.g.~$|\Psi_g\rangle = \sqrt{2}\mathrm{Re}|\Psi_+\rangle$ and $|\Psi_u\rangle =\sqrt{2}\mathrm{Im}|\Psi_+\rangle$, characterized by their parity $g/u$ under the reflection $Q_3\rightarrow -Q_3$.  In the above $M\rightarrow\infty$, $g\rightarrow 0$ limit, these  reduce to
\begin{align*}
|\Psi_g\rangle &\rightarrow \f{\rho(Q)}{\sqrt{2}} (1 + \cos\eta) |0\!\!\u 0\!\!\d g\!\!\u\rangle + \f{\rho(Q)}{\sqrt{2}} \sin\eta \, |0\!\!\u 0\!\!\d u\!\!\u\rangle \nn \\
|\Psi_u\rangle &\rightarrow \f{\rho(Q)}{\sqrt{2}} (1 - \cos\eta) |0\!\!\u 0\!\!\d u\!\!\u\rangle + \f{\rho(Q)}{\sqrt{2}} \sin\eta \, |0\!\!\u 0\!\!\d g\!\!\u\rangle {,}
\end{align*}
where $|0\!\!\u 0\!\!\d g\!\!\u\rangle  = |\phi_{0\u}\phi_{0\d}\phi_{g\u}\rangle$ with the following single particle orbitals: \begin{align}
|\phi_0\rangle &= |\phi_{k=0}\rangle  \nn \\
|\phi_g\rangle &= \sqrt{2} \mathrm{Re} |\phi_{k=+1}\rangle \nn \\
|\phi_u\rangle &= \sqrt{2} \mathrm{Im} |\phi_{k=+1}\rangle {.}
\end{align}
$|\Psi_g\rangle$ and $|\Psi_u\rangle$ are nonadiabatic electron-nuclear counterparts of the electronic states $|0\!\!\u 0\!\!\d g\!\!\u\rangle$ and $|0\!\!\u 0\!\!\d u\!\!\u\rangle$.  Our numerical results are consistent with those obtained in the two-level approximation with the electronic state constrained to the $\{|0\!\!\u 0\!\!\d g\!\!\u\rangle$, $|0\!\!\u 0\!\!\d u\!\!\u\rangle\}$ subspace.\cite{allen2005}

Symmetry analysis tells us that all of the eigenstates transform as either $A$ or $E$ irreducible representations, but only the Hamiltonian matrix elements can determine their energetic ordering and whether the ground state has $A$ or $E$ symmetry.  In the following, we choose parameters such that the ground state is degenerate ($E$ symmetry).  Two features of the model are chiefly responsible for the ground state degeneracy.  First, fermionic symmetry constrains the third electron to occupy the degenerate $k=\pm 1$ orbitals after the lowest energy $k=0$ orbital is occupied twice.  Second, electron-nuclear coupling lowers the energy of the pseudorotating $E$ states with respect to $A$ states.  
% The depth of the  cylindrically symmetrical trough in Fig.~\ref{fig:XSq:surface}c increases with increasing $g$.

\begin{figure}[t!]
\begin{tabular}{@{}cc@{}}
\includegraphics[width=0.49\columnwidth]{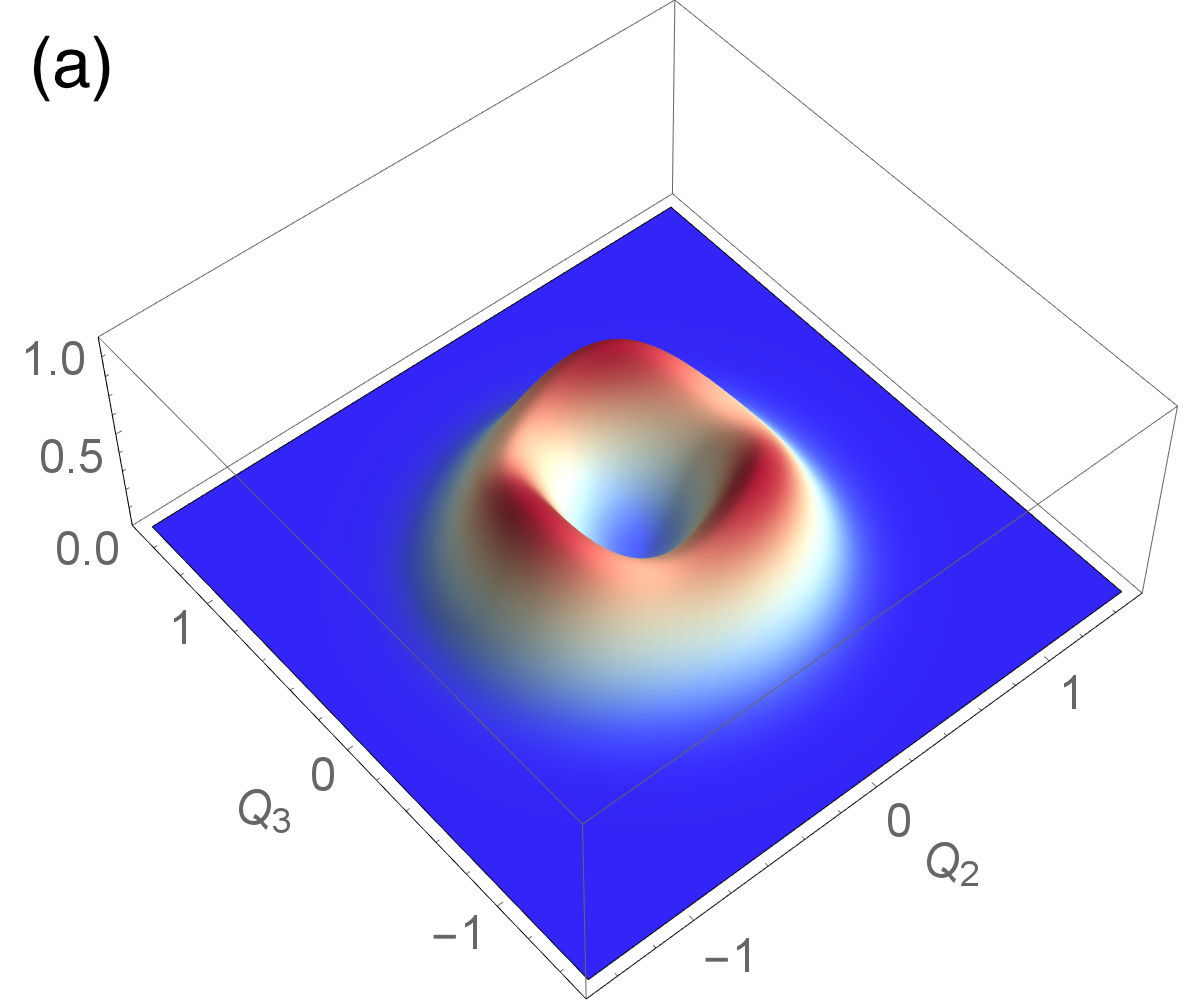} & \includegraphics[width=0.49\columnwidth]{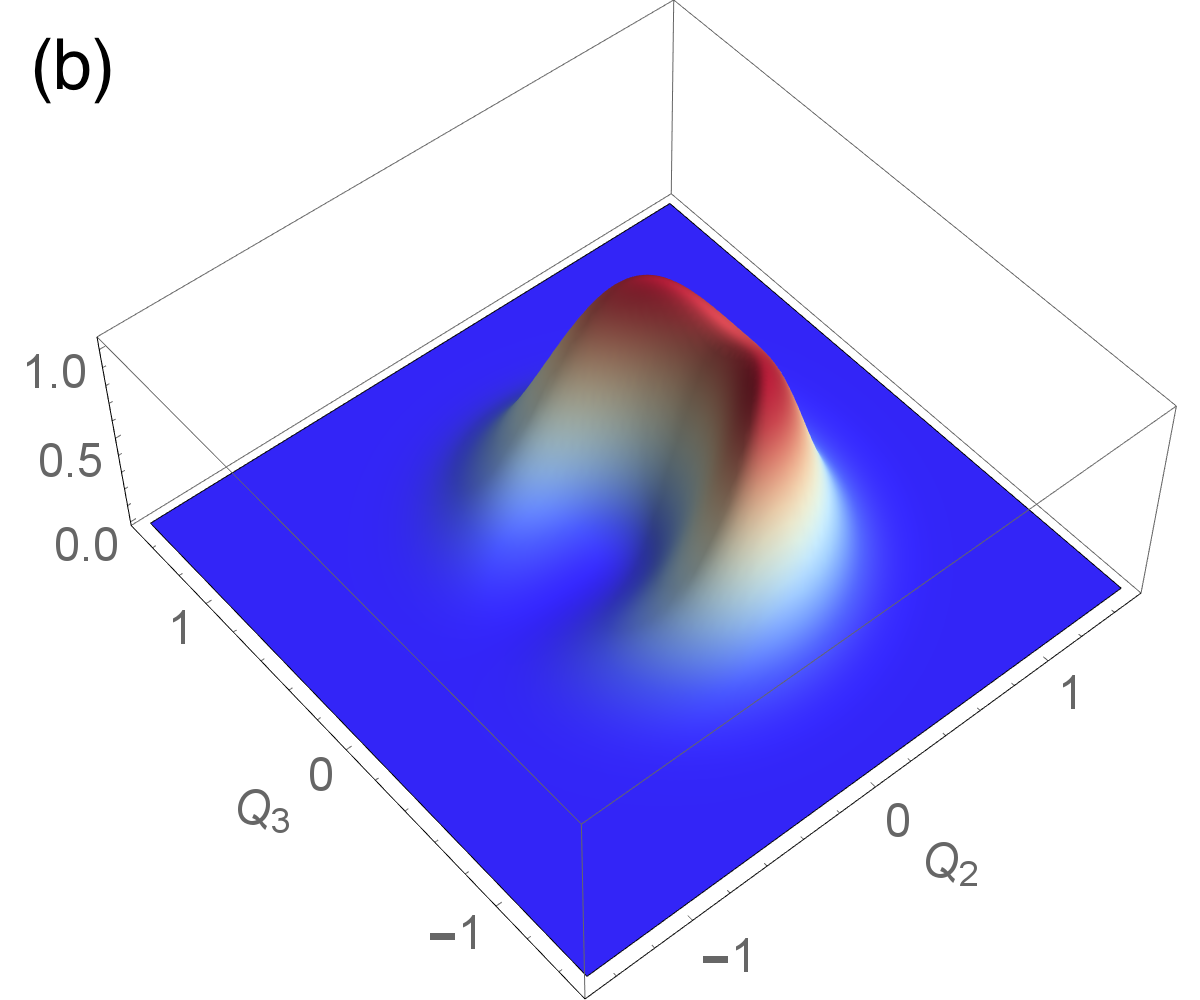} \\
\includegraphics[width=0.49\columnwidth]{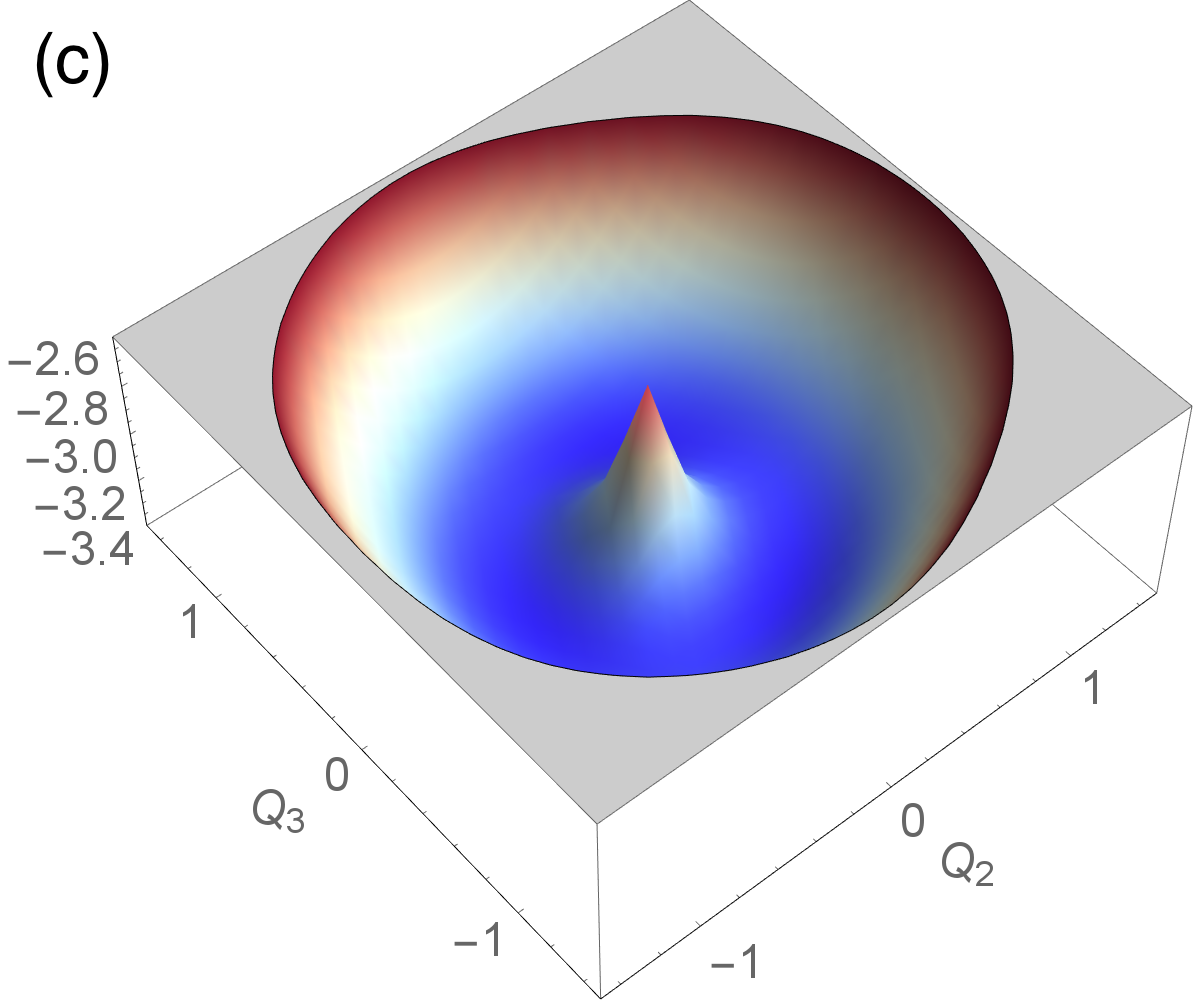} &\includegraphics[width=0.49\columnwidth]{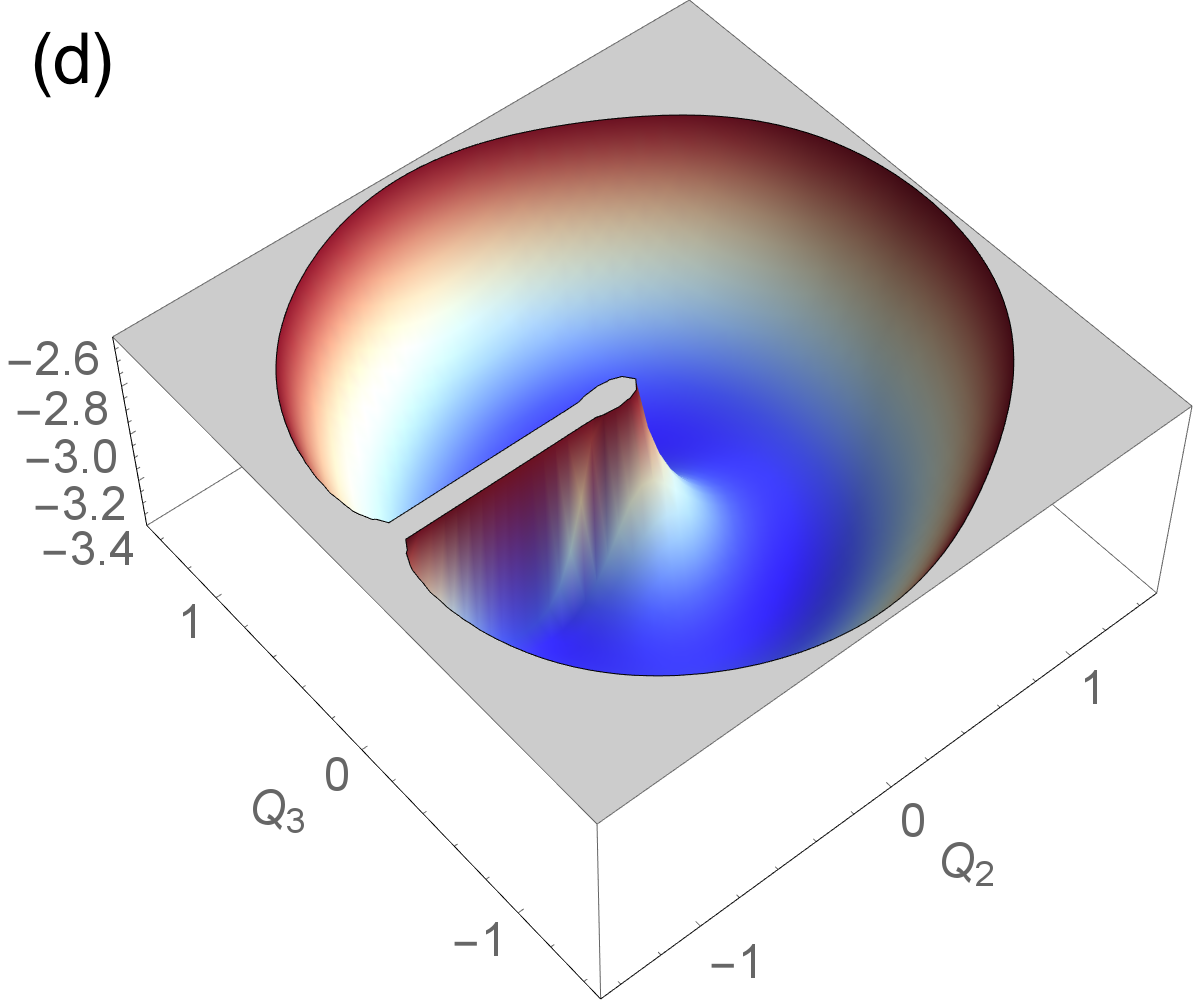} 
\end{tabular}
\caption{Nuclear probability density $|\chi|^2$ for (a) the current-carrying state $|\Psi_+\rangle$ and (b) the even parity state $|\Psi_g\rangle$.  Lower panels (c,d) show the corresponding potential energy surfaces.  Peaks in $|\chi|^2$ in (a) correspond to obtuse triangle geometries.  Model parameters are $M=24$, $K=0.6$ and $g=1.20$ and energy is measured in units of the hopping parameter $t_0$.}
\label{fig:XSq:surface} 
\end{figure}

The nuclear probability density is
\begin{align}
|\chi(R)|^2 &= \langle \Psi | \Psi\rangle_r \nn \\
&= \sum_{\alpha} |\Gamma_{\alpha}(R)|^2 {,}
\end{align}
where $\langle \cdots \rangle_r$ denotes the inner product on the electronic Hilbert space only.  In Fig.~\ref{fig:XSq:surface}, $|\chi(R)|^2$ is shown as a function of $(Q_2,Q_3)$ for model parameters $M=24$, $K=0.6$ and $g=1.2$.  The nuclear wavefunction was expanded over the basis $\langle Q\eta |nm\rangle = \f{1}{\sqrt{2\pi}} \rho_{nm}(Q) e^{im\eta}$ with $n=0,1,\ldots 13$, $m=-n,-n+2,\ldots,n$ and $|m|\leq n$.  The peaks at $\eta = (\pi/3, \pi, 5\pi/3)$ correspond to the three distinct obtuse triangle configurations, which confirms that our model is qualitatively consistent with adiabatic results for Na$_3$.\cite{delacretaz1986,coccini1988,kendrick1997,vonbusch1998}  The $E\otimes e$ Jahn-Teller model with linear electron-nuclear coupling $\propto gQ$, c.f.~Eq.~(\ref{eq:Hen:3}), has cylindrical symmetry with respect to rotations in the $(Q_2,Q_3)$ plane.  Quadratic and higher-order terms break cylindrical symmetry.  In our model, the coupling of $|a\rangle$ and $|b\rangle$ to the other six electronic states lowers the cylindrical symmetry to $C_3$ even though we have only linear electron-nuclear coupling.  The potential energy surfaces and electron-nuclear coupling of real molecules generally have significant anharmonic contributions,\cite{garcia-fernandez2005} which we are neglecting.  

Given $|\chi|^2$, the marginal nuclear probability amplitude can be written as
\begin{align}
\chi(R) = |\chi(R)| e^{\f{i}{\hbar}S(R)} {,}
\end{align}
where $S(R)$ is an arbitrary function of $R$.  Changing the phase according to $S(R)\rightarrow S(R)+\Lambda(R)$ implies a gauge transformation $\mathbf{A}_{\mu} \rightarrow \mathbf{A}_{\mu} - \nabla_{\mu}\Lambda$ of the vector potential.

\section{Molecular geometric phase \label{sec:geometricphase}}

With the exact solution in hand, we can demonstrate that the geometric phase of $\Phi_R$ is nonzero for our model pseudo\-rotating molecule.  It is instructive to review the derivation\cite{samuel1988} of the geometric phase for an arbitrary open curve $R(s)$, parameterized by $s\in [s_1,s_2]$, from $R_1$ to $R_2$.  One starts from the Pancharatnam phase $\mathrm{Arg} \langle \Phi_{R_1} | \Phi_{R_2}\rangle$, which defines a unique relative phase between the endpoints, provided $|\Phi_{R_1}\rangle$ and $|\Phi_{R_2}\rangle$ are not orthogonal.\cite{pancharatnam1956}  
The open-path geometric phase is then conventionally defined to be the remainder after subtracting the dynamical phase, $-\f{1}{\hbar}\int_{s_1}^{s_2} \langle \Phi_{R(s)} | \hat{H}(s) | \Phi_{R(s)} \rangle ds$, from the Pancharatnam phase.\cite{mukunda1993}  Here, $\hat{H}(t)$ is an auxiliary Hamiltonian that drives the electronic wavefunction along the path $|\Phi_{R(s)}\rangle$ in Hilbert space, i.e.~$|\Phi_{R(s)}\rangle$ is the solution of the Schr\"odinger equation $i\hbar\partial_s |\Phi\rangle = \hat{H}(s) |\Phi\rangle$.  Since our definition $\mathbf{A}_{\mu}=\hbar\:\mathrm{Im}\langle \Phi_R | \nabla_{\mu} \Phi_R\rangle$ differs by a sign from the conventional definition $\mathbf{A}_{\mu}=i\hbar\langle \Phi_R | \nabla_{\mu} \Phi_R\rangle$, we have introduced a minus sign in the following definition of the exact molecular geometric phase:
\begin{align}
\gamma &= -\mathrm{Arg} \langle \Phi_{R_1} | \Phi_{R_2}\rangle - \f{1}{\hbar}\int_{s_1}^{s_2} \langle \Phi_{R(s)} | \hat{H}(s) | \Phi_{R(s)} \rangle ds \nonumber \\
&= -\mathrm{Arg} \langle \Phi_{R_1} | \Phi_{R_2}\rangle + \f{1}{\hbar}\int_{R_1}^{R_2} \mathbf{A}_{\mu}\cdot d\R_{\mu} 
\label{eq:geometricphase}
\end{align}
with an implicit sum over nuclei $\mu$.  Equation~(\ref{eq:geometricphase}) extends the familiar geometric phase to open paths in a way that maintains gauge invariance under $R$-dependent gauge transformations of $|\Phi_R\rangle$.  Like the Aharonov-Anandan phase,\cite{aharonov1987} it does not assume an adiabatic approximation.

Equation~(\ref{eq:geometricphase}) is also valid in the Born-Oppenheimer approximation.  In that approximation, there are two choices of gauge for which the Longuet-Higgins phase $\pi$ can be simply understood.\cite{sjoqvist1998}  First, if the phase of $\Phi_R(r)$ is chosen such that $\Phi_R(r)$ is real-valued for all $R$, then $\mathbf{A}_{\mu}=0$ and the second term of Eq.~(\ref{eq:geometricphase}) vanishes.  Hence, the phase $\pi$ comes from the first term and the sign change of $\Phi_R(r)$, i.e.~$\Phi_{R_2}(r)=-\Phi_{R_1}(r)$ (multivaluedness).  Alternatively, if $\Phi_R(r)$ is chosen to be single-valued, then the first term vanishes for $R_1=R_2$ and the second term gives the phase $\pi$ because there is no gauge for which the vector potential is zero everywhere along the path.  Applying $R$-dependent gauge transformations to $\Phi_R(r)$ and $\chi(R)$ changes the first and second terms of Eq.~(\ref{eq:geometricphase}), but their sum remains the same. 

Using the definition of the conditional electronic wavefunction $\Phi_R(r) = \Psi(r,R)/\chi(R)$, the vector potential can be expressed as 
\begin{align}
\mathbf{A}_{\mu} &= \f{\hbar\,\mathrm{Im} \langle \Psi | \nabla_{\mu} \Psi \rangle_r}{|\chi|^2} - \nabla_{\mu} S {.}
\label{eq:A}
\end{align}
The first term of Eq.~(\ref{eq:A}) is gauge invariant.  It is responsible for the geometric phase when the exact factorization is applied to our model system.

Since all three nuclear coordinates $\R_{\mu}$ are determined by $(Q,\eta)$, it is convenient to define the vectors $\R = Q \mathbf{e}_Q$ and $\mathbf{A} = A_Q \mathbf{e}_Q + A_{\eta} \mathbf{e}_{\eta}$ in order to simplify the notation.  For a closed path $\mathcal{C}$, the geometric phase is 
\begin{align}
\gamma = \f{1}{\hbar} \oint_{\mathcal{C}} \mathbf{A}\cdot d\R = \f{1}{\hbar} \oint_{\mathcal{C}} \big( A_{Q} dQ + A_{\eta} Q d\eta \big) {,}
\end{align}
where $A_Q = \hbar\:\mathrm{Im} \langle \Phi_R | \partial_Q \Phi_R \rangle$; $A_{\eta} =  (\hbar/Q)\mathrm{Im} \langle \Phi_R | \partial_{\eta} \Phi_R \rangle$.

\begin{figure}[t!]
\includegraphics[width=0.95\columnwidth]{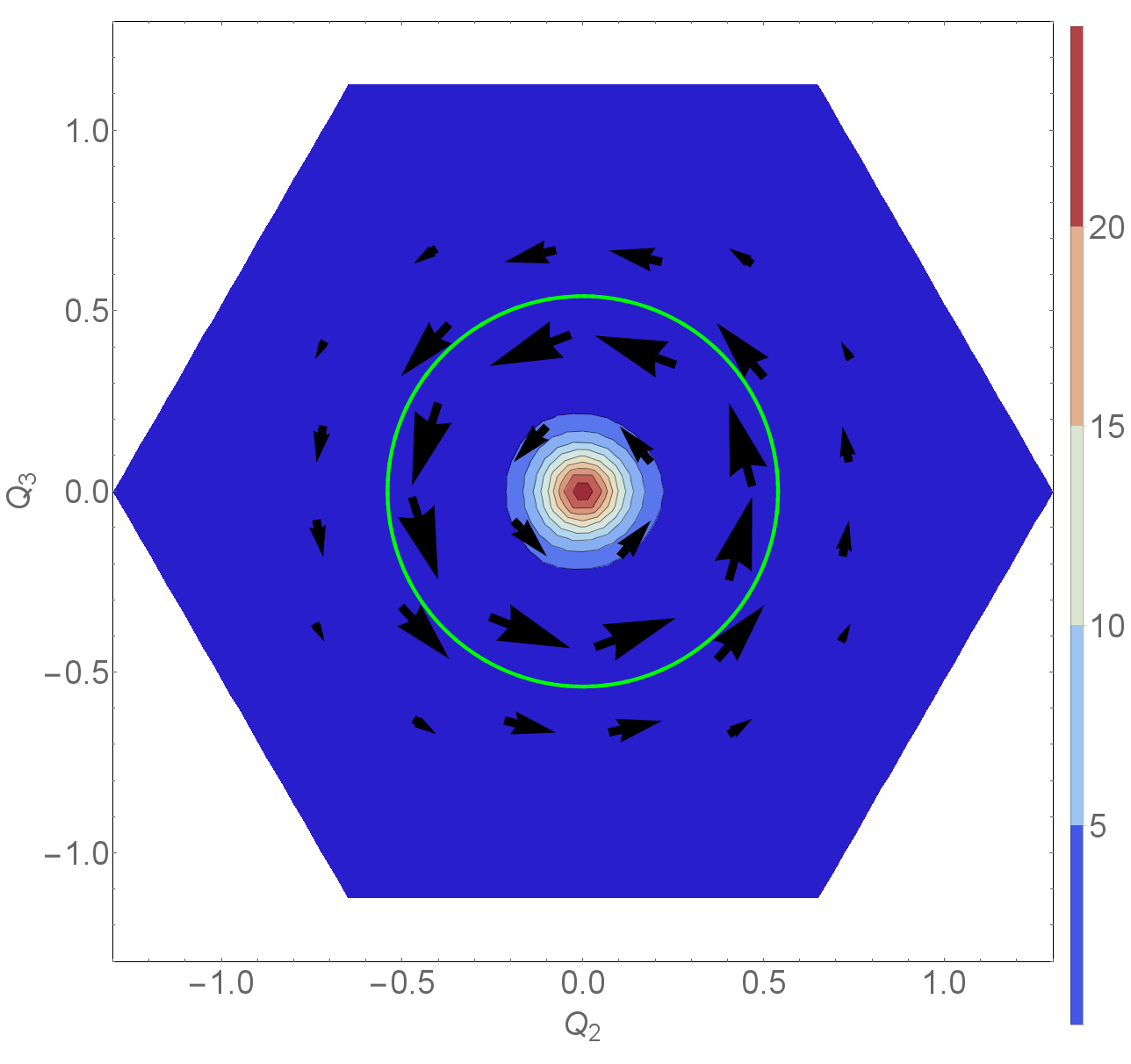}
\caption{Nuclear current vector field (arrows) superimposed on a color scale plot of the Berry curvature $B_{Q_2Q_3}$ for $|\Psi_+\rangle$.  The green circle shows the path on which the geometric phase (Fig.~\ref{fig:phase}a) and various cross-sections (Fig.~\ref{fig:warping}) are calculated.}
\label{fig:curvature:current} 
\end{figure}

\begin{figure}[b!]
\begin{tabular}{@{}cc@{}}
\includegraphics[width=0.48\columnwidth]{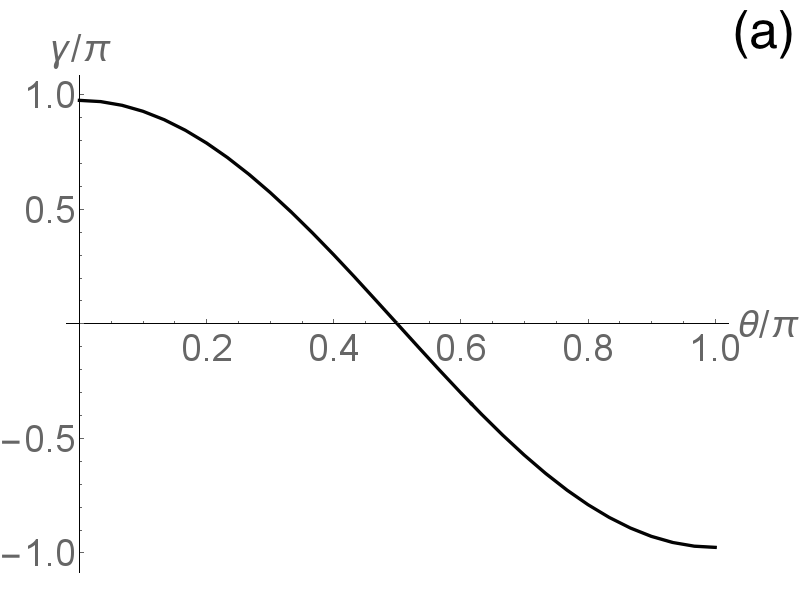} &
\includegraphics[width=0.48\columnwidth]{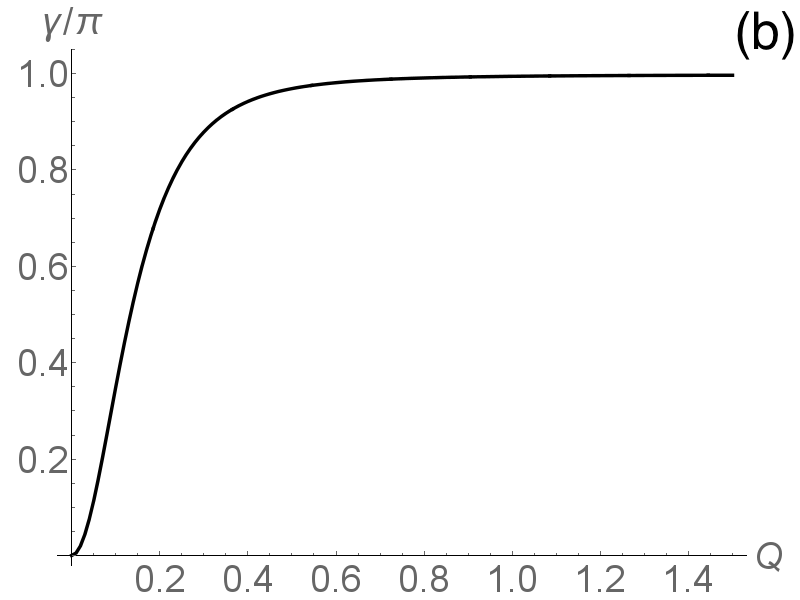} 
\end{tabular}
\caption{Molecular geometric phase (a) calculated along the circular path in Fig.~\ref{fig:curvature:current} versus the mixing angle $\theta$ in Eq.~(\ref{eq:psi:mixed}) (b) as a function of path radius $Q$ for $\theta=0$.}
\label{fig:phase} 
\end{figure}

Ground state degeneracy has important consequences for the molecular geometric phase and vector potential.  Using the parameters $(\theta,\varphi)$ to express an arbitrary state in the degenerate ground state manifold, we have
\begin{align}
|\Psi\rangle = \cos\f{\theta}{2} e^{-i\varphi/2} |\Psi_+\rangle + \sin\f{\theta}{2} e^{i\varphi/2} |\Psi_-\rangle {.}
\label{eq:psi:mixed}
\end{align}
When $\theta=0$, the state has positive (counterclockwise) nuclear current in the $(Q_2,Q_3)$ plane as shown in Fig.~\ref{fig:curvature:current}.
For $\theta=\pi$, the current is equal and opposite.  Thus, $\theta$ tunes the current continuously between its maximum values at the poles of the Bloch sphere.\cite{allen2005}  Figure~\ref{fig:phase}a shows the geometric phase calculated along the circular path $\{Q=0.54, \;\eta=[0,2\pi]\}$ as a function of $\theta$.  Unlike the Longuet-Higgins phase, the exact geometric phase is not quantized to 0 or $\pi$ and varies in proportion to the nuclear current carried by the state.  The exact factorization scheme can also be applied to adiabatic electron-nuclear eigenstates such as the limiting function in Eq.~(\ref{eq:limit}).  In this case as well, the geometric phase can be tuned by forming superpositions of degenerate states analogously to Eq.~(\ref{eq:psi:mixed}).  Nevertheless, the geometric phase of such a state, $\gamma^{BO} = \pi \cos\theta$, is still a topological quantity, which can only take the values $n\gamma^{BO}$ with integer $n$.

For $\theta=\pi/2$, $|\Psi\rangle$ is real and the current and geometric phase vanish for any value of $\varphi$, since it is always possible to choose a gauge such that $\mathbf{A}$ is identically zero.  
However, $C_3$ symmetry is broken and the nuclear probability density is displaced in the $\eta=\varphi$ direction of the $(Q_2,Q_3)$ plane.  
% For any choice of $\varphi$, $C_3$ symmetry is broken, displacing the nuclear probability in the $\eta=\varphi$ direction of the $(Q_2,Q_3)$ plane.  
For example, $\varphi=0$ gives $|\Psi_g\rangle = \sqrt{2}\mathrm{Re} |\Psi_+\rangle$ for which $|\chi(R)|^2$ is offset in the $\eta=0$ direction; cf.~Fig.~\ref{fig:XSq:surface}b.  The value $\varphi=\pi$ gives the odd state $|\Psi_u\rangle$. 

The topological Longuet-Higgins phase depends only on whether the path $\mathcal{C}$ encloses a conical intersection, otherwise it is path independent.  This is not the case for the exact molecular geometric phase.  Figure~\ref{fig:phase}b shows the value of the geometric phase calculated on a circular path as a function of the radius $Q$.  The geometric phase is path dependent because via Stokes' theorem it depends on the net flux of the effective magnetic field $\mathbf{B} = \nabla\times\mathbf{A}$ through the surface $\mathcal{S}$ bounded by $\mathcal{C}$:
\begin{align}
\gamma &= \f{1}{\hbar} \iint_{\mathcal{S}} \mathbf{B}\cdot d\mathbf{S} \nn \\
&= \f{1}{\hbar} \iint_{\mathcal{S}} B_{\mu\nu} dq^{\mu} \wedge dq^{\nu}
\label{eq:flux}
\end{align}
where the antisymmetric wedge product is used to write a general expression in terms of arbitrary coordinates $q^{\mu}$.\cite{berry1989}  The Berry curvature
\begin{align}
B_{\mu\nu} &= \hbar\:\mathrm{Im} \langle \partial_{\mu} \Phi_R | \partial_{\nu} \Phi_R \rangle 
\end{align}
is related to the field strength by $B_{Q_1} = \f{1}{Q} \epsilon_{\mu\nu z} B_{\mu\nu}$ with $q^{1}=Q_2$ and $q^2=Q_3$.  Since $B_{\mu\nu}$ is an antisymmetric tensor it has only one independent element $B_{Q_2 Q_3}$ which is plotted in Fig.~\ref{fig:curvature:current}.  For $|\Psi_+\rangle$ in the $M\rightarrow\infty$ limit, the magnetic flux becomes localized at the origin in the $(Q_2,Q_3)$ plane and approaches $\f{h}{2}\delta(Q_2)\delta(Q_3)$, thereby recovering the well known adiabatic result.  In the three-dimensional nuclear coordinate space $(Q_1,Q_2,Q_3)$, the flux coincides with the line (generally, a submanifold of codimension 2), parametrized by $Q_1$, on which the two adiabatic potential energy surfaces undergo a conical intersection.  The magnetic field is thus equivalent to that of an infinitesimal flux tube carrying flux $h/2$ and its Aharonov-Bohm phase shift is responsible for the discrete Longuet-Higgins phase $\pi$.  In the $(Q_2,Q_3)$ plane, the vector potential of the flux tube is equivalent to that of a magnetic monopole with charge $g=1/2$ at the origin.

\begin{figure}[t!]
\includegraphics[width=0.48\columnwidth]{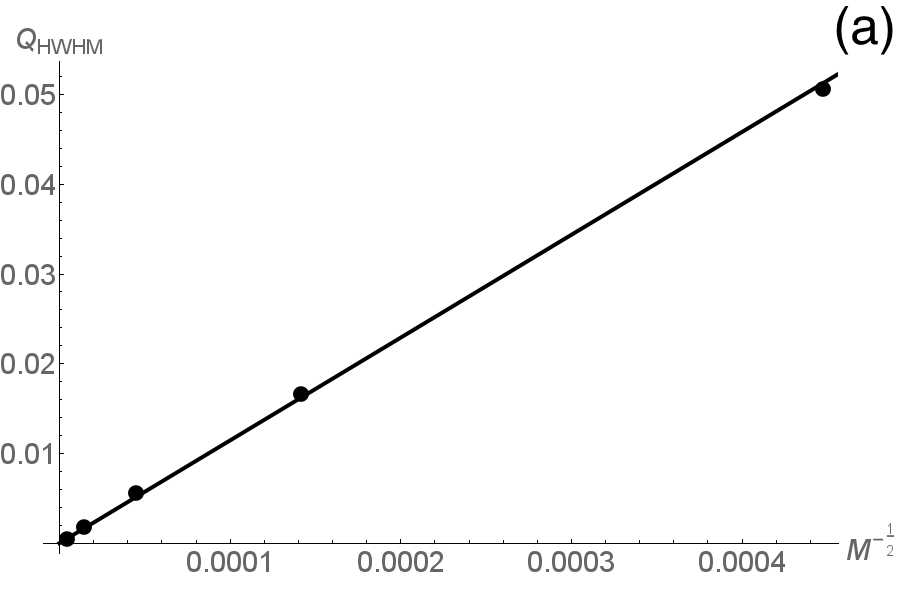}
\includegraphics[width=0.48\columnwidth]{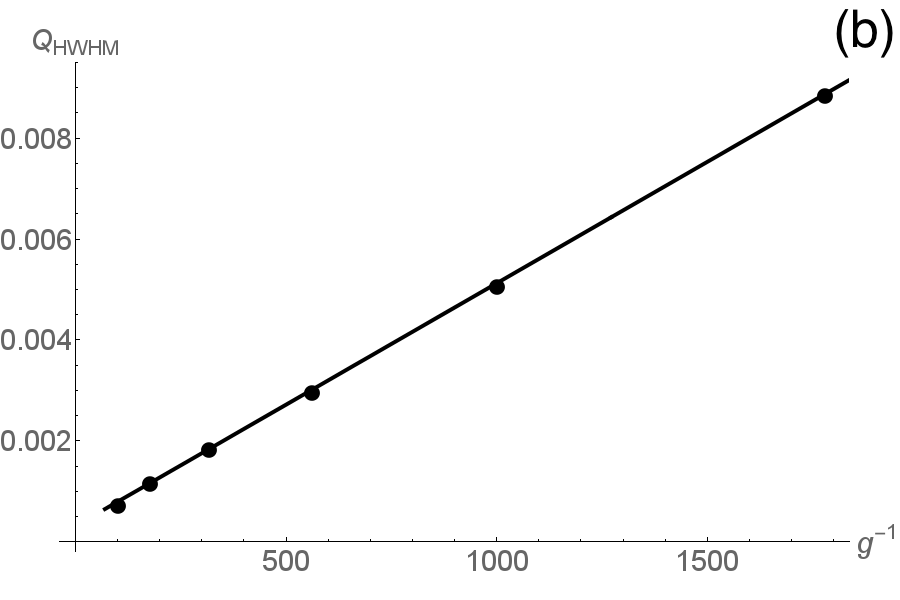}
\caption{Dependence of the characteristic radius $Q_{\rm HWHM}$ on $M$ and $g$.  Calculations were performed for (a) $K=2/9$ and $g=0.01$ and (b) $K=2/9$ and $M=10^{10}$.}
\label{fig:fits} 
\end{figure}

In the exact factorization, the flux tube (or monopole) gets smeared out over a finite area, i.e.~the point-like flux is replaced by an extended flux density $\mathcal{B}(Q_1,Q_2,Q_3)$ that satisfies $\iint \mathcal{B}(Q_1,Q_2,Q_3) dQ_2 dQ_3 = h/2$.  Unlike the adiabatic case, the geometric phase vanishes if the path encircling the origin is shrunk to a point.  On the other hand, if the radius of the path is taken to infinity, the geometric phase approaches $\pi$, as shown in Fig.~\ref{fig:phase}b, since then all of the flux is enclosed.  Similarly, the geometric phase calculated on a path with any finite radius $Q$ approaches $\pi$ in the limit $M\rightarrow\infty$, since the characteristic radius of the flux tube tends to 0.  The characteristic radius can be defined as the half width at half maximum of the peak in $B_{Q_2Q_3}$ at the origin.  Figure~\ref{fig:fits}a shows that $Q_{\rm HWHM}$ decreases as $M^{-1/2}$ as $M\rightarrow \infty$, consistent with the result $Q_{\rm HWHM} \sim \f{\hbar K^{1/2}}{g M^{1/2}}$ of an asymptotic analysis to be reported elsewhere.  Figure~\ref{fig:fits}b shows that $Q_{\rm HWHM}$ varies as $g^{-1}$ if $M$ is sufficiently large.

For a complex Hamiltonian with three-dimensional slow parameter space $(X,Y,Z)$, such as Berry's original two-level example,\cite{berry1984} degeneracies of the adiabatic potential energy surfaces occur at isolated points and the geometric phase can be calculated as the flux of a magnetic monopole with charge $g=1/2$ located at the point of degeneracy.  If the total Hamiltonian includes a kinetic energy operator for the slow variables and the separation of fast and slow variables is made using the exact factorization instead of the Born-Oppenheimer approximation, then the geometric phase can be calculated in the same way but the point monopole will be replaced by a smeared-out magnetic charge density $\rho(X,Y,Z)$ which integrates to $1/2$.

The nuclear current density of the state $|\Psi\rangle$ is
\begin{align}
\mathbf{J} &= \mathcal{M}^{-1} \hbar\,\mathrm{Im} \langle \Psi | \nabla \Psi\rangle_r \nn \\
&=\mathcal{M}^{-1} \big( \hbar\, \mathrm{Im} \chi^* \nabla \chi + \mathbf{A} |\chi|^2 \big) {.}
\end{align}
For the state $|\Psi_+\rangle$, the nuclear current circulates in the positive direction around the origin of the $(Q_2,Q_3)$ plane, see Fig.~\ref{fig:curvature:current}, while the electronic \textit{particle} current (not to be confused with the \textit{charge} current) circulates around the three-site ring in the positive direction $0\rightarrow 1 \rightarrow 2 \rightarrow 0$. In our model, the electronic circulating current is 
\begin{align}
\hat{J}_e = \sum_{n\s} \big( -i t_{n,n+1} c_{n\s}^{\dag} c_{n+1\s} + H.c.\big) {.} \label{eq:current:elec}
\end{align}
Looking at the spin-resolved electronic current
\begin{align}
\hat{J}_{e\s} = \sum_{n} \big( -i t_{n,n+1} c_{n\s}^{\dag} c_{n+1\s} + H.c.\big) 
\end{align}
reveals that spin up electrons circulate in the positive direction, while somewhat fewer spin down electrons circulate in the negative direction. 

To summarize, when the Born-Oppenheimer product is replaced by the exact factorization, the molecular geometric phase is no longer quantized and can take any value between 0 and $2\pi$ even though the Hamiltonian is real valued.  
\begin{figure}[b!]
\hspace{0.24cm}\includegraphics[width=0.85\columnwidth]{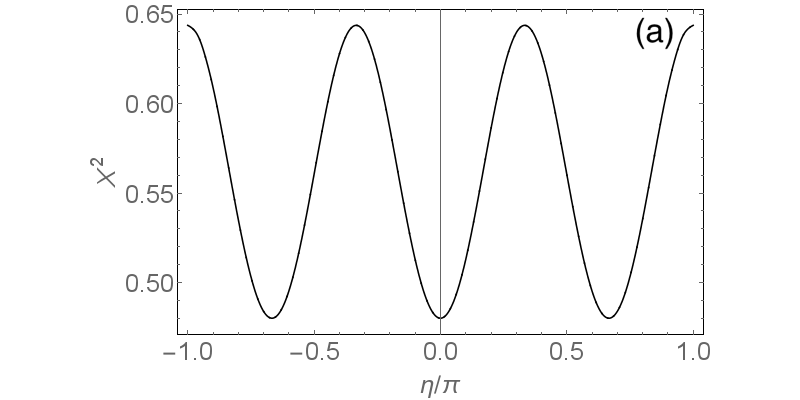} \\\includegraphics[width=0.85\columnwidth]{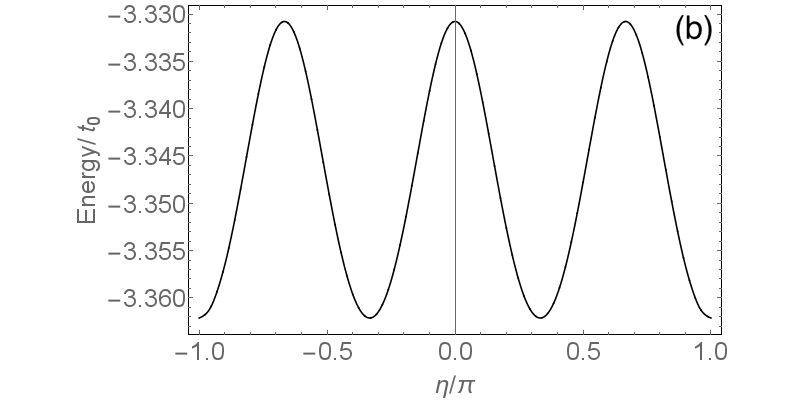} 
\caption{One-dimensional circular slice ($Q=0.54$, $\eta=[0,2\pi]$) of (a) $|\chi|^2$ in Fig.~\ref{fig:XSq:surface}a and (b) $\mathcal{E}(R)$ in Fig.~\ref{fig:XSq:surface}c.}
\label{fig:warping} 
\end{figure} 
Nonzero geometric phase implies that $\mathbf{A}_{\mu}$ cannot be gauged away and hence gives a nontrivial contribution to the nuclear current density.  The curl of $\mathbf{A}_{\mu}$ is an induced magnetic field, which can be viewed as the field a smeared-out Aharonov-Bohm flux tube.  Spin-orbit interactions cause a similar spreading in the adiabatic case.\cite{mead1980b}  
% Spin-orbit interactions cause a similar spreading as has been shown in the adiabatic case.\cite{mead1980b}
However, the smearing effect we have studied here is different because it is a nonadiabatic effect of the nuclear kinetic energy operator that occurs even when spin-orbit interactions are neglected.

Since our model takes into account the full eight-dimensional basis of three-electron states as opposed to only the two lowest energy states of the $E\otimes e$ Jahn-Teller model, electron-electron interactions can be represented in the model Hamiltonian.  We have performed calculations with a Hubbard term $U(n_{1\u}n_{1\d} +n_{2\u}n_{2\d} + n_{3\u}n_{3\d})$ and verified that it does not change our main results. 

We have neglected nuclear exchange symmetry and rotations, and therefore the question arises whether the molecular geometric phase will survive if they are included in the problem.  Since the occurrence of nontrivial molecular geometric phase depends critically on the degeneracy of the state, we end this section by commenting on the implications of some known degeneracies of the general electron-nuclear problem for the molecular geometric phase and induced vector potential.  
\begin{figure}[t!]
\includegraphics[width=0.85\columnwidth]{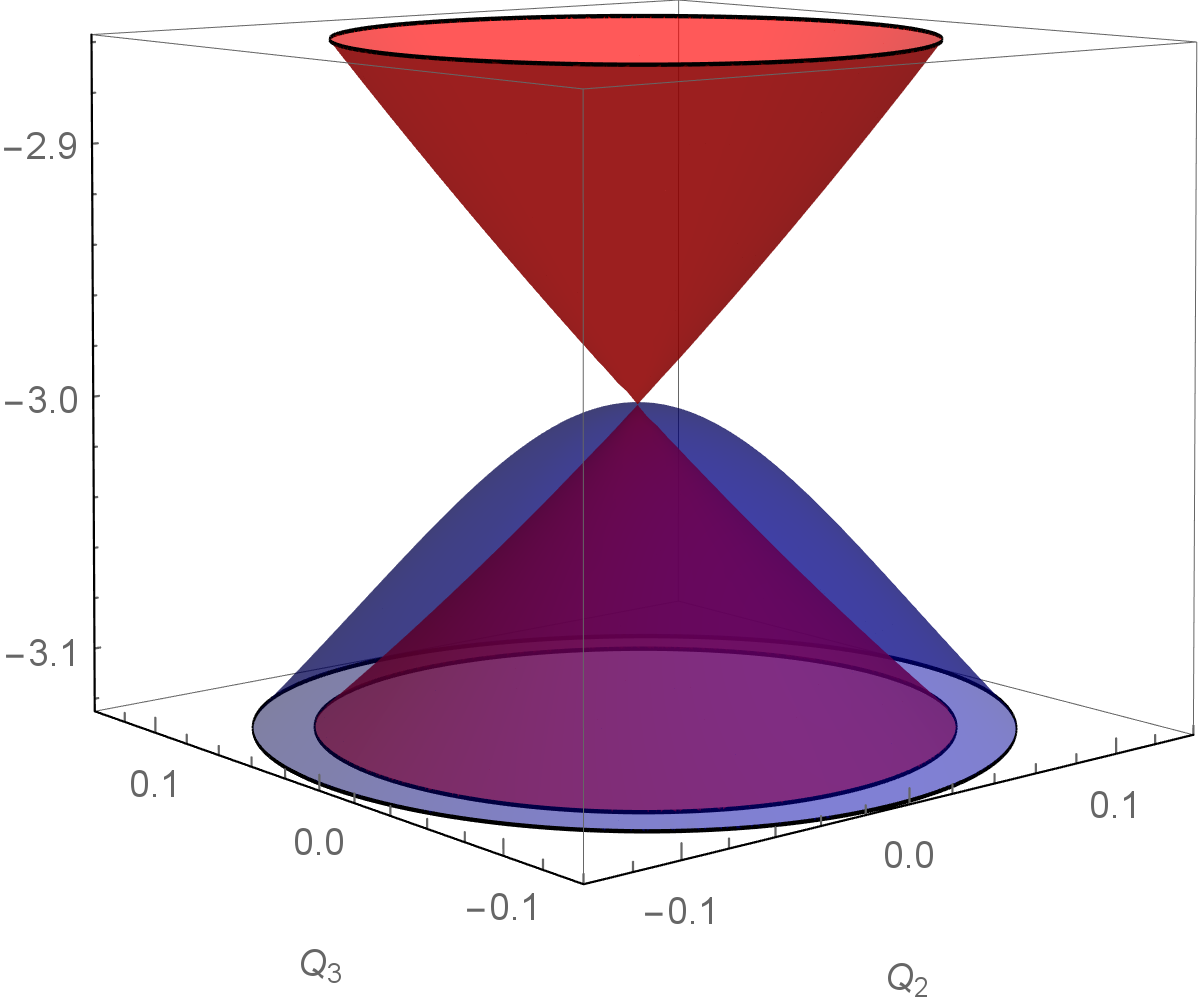} 
\caption{Born-Oppenheimer energy surface $\langle \Phi_R^{BO} | \hat{H}^{BO} | \Phi_R^{BO} \rangle$ (red) and the potential term $\langle \Phi_R | \hat{H}^{BO} | \Phi_R \rangle$ (blue) for the state $|\Psi_+\rangle$.  The exact potential energy surface is the sum of $\langle \Phi_R | \hat{H}^{BO} | \Phi_R \rangle$ and the metric term in Fig.~\ref{fig:metric}.}
\label{fig:adiabatic} 
\end{figure} 
The energy eigenstates of any system of electrons and nuclei can be chosen to be simultaneous eigenstates of the total angular momentum operators $\hat{J}^2,\hat{J}_z$, due to the isotropy of space.  The resulting $2J+1$ degeneracies are vital for the exact molecular geometric phase because they make it possible to construct current-carrying eigenstates from complex superpositions of degenerate states.  If the molecular geometric phase of a current-carrying state is nonzero, the nuclear Schr\"odinger equation must contain induced vector potentials which contribute to the nuclear current.  Whether the molecular geometric phase is appreciably different from its adiabatic value in a particular system depends on a number of factors including the strength of the electron-nuclear coupling (represented by $g$ in our model) and the rigidity of the molecule.  Systems in which the electronic state is nonadiabatically excited by large amplitude nuclear motions and floppy molecules, such as pseudorotating molecules or highly excited molecules, are more likely to have a molecular geometric phase that differs significantly from the adiabatic value.

\section{Potential energy surface \label{sec:PES}}

The Schr\"odinger equation for the nuclei in the exact factorization has the same form as Eq.~(\ref{eq:BO:mead}) except the adiabatic potential energy surface $\mathcal{E}^{BO}$ and vector potential $\mathbf{A}_{\mu}^{BO}$ are replaced by their exact counterparts.  The exact potential energy surface can be expressed as\cite{gidopoulos2014}    
\begin{align}
\mathcal{E}(R) &= \langle \Phi_R | \hat{H}^{BO} | \Phi_R \rangle \nn\\
&+ \sum_{\mu} \f{\hbar^2 \langle \nabla_{\mu} \Phi_R | \nabla_{\mu} \Phi_R \rangle}{2M_{\mu}} - \sum_{\mu} \f{|\mathbf{A}_{\mu}|^2}{2M_{\mu}}
\label{eq:surface}
\end{align}
where $\hat{H}^{BO} = \hat{H} - \hat{T}_n$.  Equation~\ref{eq:surface} is analogous to the expression for the adiabatic potential energy surface including the Born-Huang nuclear gradient terms.\cite{born1954,zygelman1987}

The potential energy surface for the nuclear factor $\chi_+$ of the current-carrying state $|\Psi_+\rangle$ is plotted in Fig.~\ref{fig:XSq:surface}c.  It appears to have rotational symmetry with respect to the pseudorotational angle $\eta$, however the one-dimensional cut along the circle $Q=0.54$ in Fig.~\ref{fig:warping} reveals a weak threefold symmetric warping consistent with $|\chi_+|^2$.  

The potential energy surface for the nuclear factor $\chi_g$ in Fig.~\ref{fig:XSq:surface}d has a high and narrow barrier, which is cut off by our choice of scale.  As mentioned in the introduction, this barrier correlates with the $C_3$ symmetry breaking of $|\Psi_g\rangle$.  The potential energy surface shows such a barrier for any real-valued $|\Psi\rangle$, and its direction is determined by the angle $\varphi$ in Eq.~(\ref{eq:psi:mixed}).
Similar to what was found for diatomic molecules\cite{czub1978,hunter1980,hunter1981,cassam-chenai2006,lefebvre2015} and a two-mode vibronic model,\cite{chiang2014} the barrier in $\mathcal{E}(R)$ corresponds to nuclear configurations where $|\chi(R)|^2$ drops close to zero.  

The exact potential energy surface is compared with the Born-Oppenheimer surface in Fig.~\ref{fig:adiabatic}.  The cusp of the Born-Oppenheimer surface due to the conical intersection at the origin gets smoothed out in the exact surface.

\section{Quantum geometric tensor \label{sec:QGT}}

The last two terms of Eq.~(\ref{eq:surface}) can be combined into a single quantity 
\begin{align}
\mathcal{E}_{\rm geo} = \f{\hbar^2}{2}  Q^{\mu\nu} g_{\mu\nu} {,}
\label{eq:metric:term}
\end{align}
where $Q^{\mu\nu}$ is the inverse inertia tensor appearing in the nuclear kinetic energy $\f{1}{2} Q^{\mu\nu} P_{\mu}P_{\nu}$ and $g_{\mu\nu}$ is a Riemannian metric (Fubini-Study metric) defined as \cite{provost1980}
\begin{align}
g_{\mu\nu} = \mathrm{Re} \langle \nabla_{\mu} \Phi_R | 1- |\Phi_R\rangle\langle\Phi_R|| \nabla_{\nu} \Phi_R \rangle {.}
\end{align}
This is directly analogous to the result already known in the Born-Oppenheimer approximation.\cite{berry1989,shapere1989,berry1990,berry1993,goldhaber2005}
The gradient of $\mathcal{E}_{\rm geo}$ gives a geometric contribution to the electric field acting on the nuclei.\cite{berry1989,berry1990,berry1993}
The metric $g_{\mu\nu}$ and Berry curvature $B_{\mu\nu}$ can be unified into a quantum geometric tensor.\cite{berry1989}  
\begin{figure}[h!]
\includegraphics[width=0.8\columnwidth]{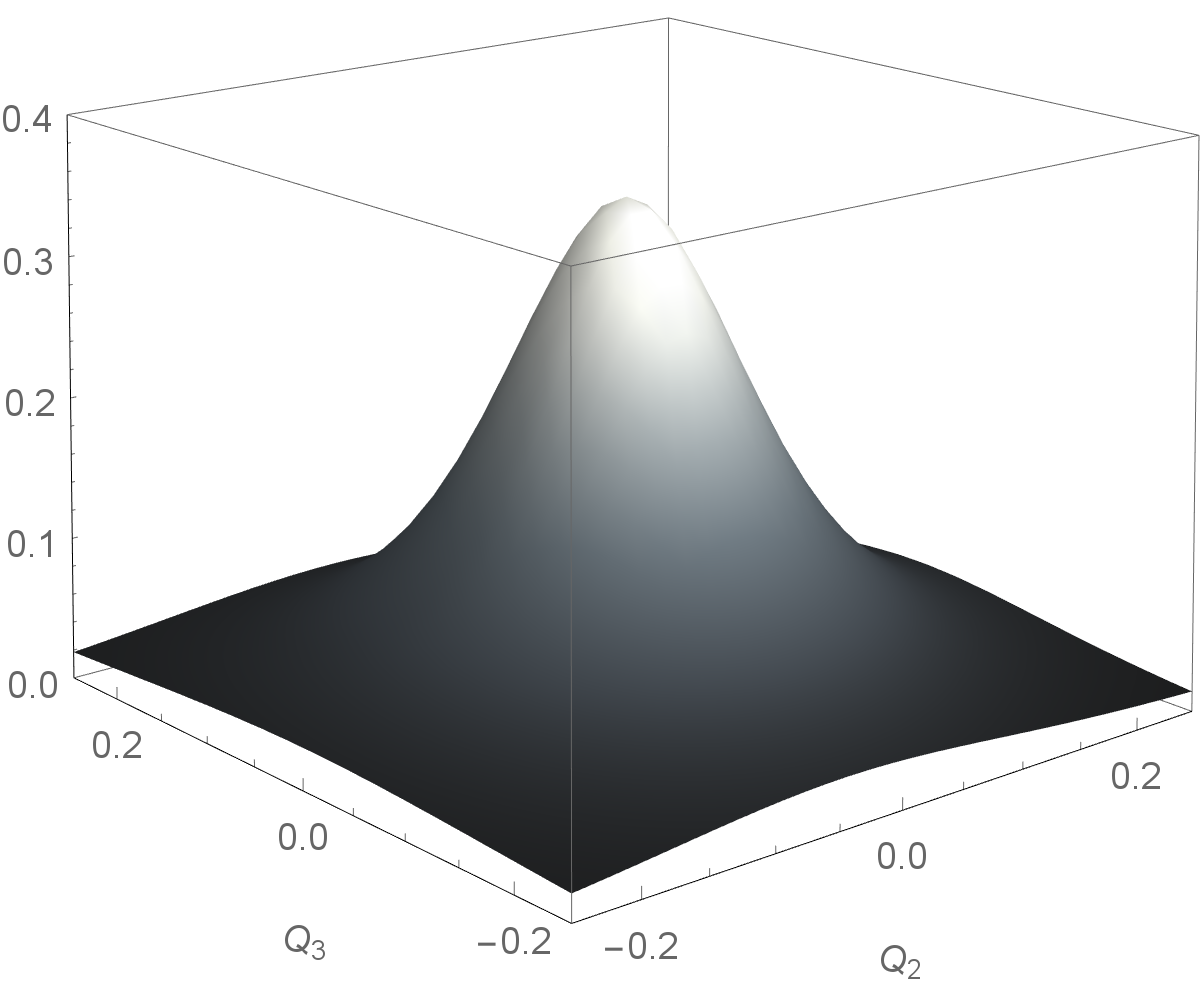} 
\caption{Geometric contribution $\mathcal{E}_{\rm geo}$ to the potential energy.}
\label{fig:metric} 
\end{figure} 
Switching to arbitary collective nuclear coordinates $q^{\mu}$, the quantum geometric tensor is
\begin{align}
T_{\mu\nu} = \langle \partial_{\mu} \Phi_R | 1- |\Phi_R\rangle\langle\Phi_R|| \partial_{\nu} \Phi_R \rangle {.}
\end{align}
The real part of $T_{\mu\nu}$ is the metric $g_{\mu\nu}$ while the imaginary part is $1/\hbar$ times the Berry curvature $B_{\mu\nu}$.  

In the adiabatic approximation, the potential $\mathcal{E}_{\rm geo}$ diverges with the inverse square distance from a conical intersection between two energy surfaces.\cite{berry1989,berry1993}  By repelling the nuclei from the vicinity of the conical intersection where the adiabatic approximation breaks down, the potential $\mathcal{E}_{\rm geo}$ enhances the accuracy of that approximation.\cite{berry1989}
In line with what we found for the vector potential and potential energy surface in previous sections, the singularity of the adiabatic  $\mathcal{E}_{\rm geo}^{BO}$ is smoothed out in the exact quantity.  Figure~\ref{fig:metric} shows that the divergent adiabatic potential is rounded off to a smooth finite peak when the metric $g_{\mu\nu}$ is evaluated with the exact electronic function $\Phi_R$ instead of the Born-Oppenheimer function $\Phi_R^{BO}$.

\section{Conclusions \label{sec:conclusions}}

The topological Longuet-Higgins phase \cite{herzberg1963} and accompanying vector potential \cite{mead1979} are observable in numerous experiments, e.g.~spectroscopy of triatomic molecules \cite{delacretaz1986,coccini1988,vonbusch1998} and 
dynamical Jahn-Teller defects in bulk crystals \cite{englman1972,ham1972,ham1987}.  The identification of such topological phases has always been based on the Born-Oppenheimer approximation, leaving open the possibility that their topological character is an artifact of that approximation which would not survive in an exact calculation.  By identifying a specific case where a nontrivial Longuet-Higgins phase of $\pi$ does indeed vanish in an exact calculation based on the exact electron-nuclear factorization, recent work has amplified this uncertainty.\cite{min2014}  In the model of Ref.~\onlinecite{min2014}, the molecular geometric phase only takes the adiabatic value $\pi$ if there is a cusp (nonanalyticity) in the potential energy surface.
% the molecular geometric phase is only nonzero if there is a cusp (nonanalyticity) in the potential energy surface.  
Since the exact potential energy surface is smooth and a cusp is only recovered in the limit $M\rightarrow \infty$, the molecular geo\-metric phase jumps discontinuously from $\pi$ to 0 when the mass is decreased from infinity to a large finite value, i.e.~when the nuclear kinetic energy is turned on.

However, it would be puzzling if that behavior were to occur in the classical models of pseudorotating molecules because in those models the Longuet-Higgins phase is a topological invariant identifiable from qualitative global properties of the electronic Born-Oppenheimer wavefunction $\Phi_R^{BO}(r)$ far away from the conical intersection\cite{longuet-higgins1975} and therefore robust to perturbations.  For large $M$, the perturbation induced by turning on the kinetic energy is localized near the conical intersection and should not affect its global properties.  Therefore, the exact $\Phi_R(r)$ is almost everywhere similar to $\Phi_R^{BO}(r)$ and the geometric phase should not be expected to jump from $\pi$ to 0.

To resolve this discrepancy, we have applied the exact factorization to a model pseudorotating molecule which is closer in spirit to the original example of Herzberg and Longuet-Higgins.\cite{herzberg1963}  We have been able to answer the two questions raised in the introduction.  First, when the Born-Oppenheimer factorization is replaced by the exact factorization, the conical intersection of the adiabatic potential energy surfaces is smoothed out, and second, the Longuet-Higgins phase becomes a path-dependent $U(1)$ geometric phase.  
This is our main result: quantities that were discrete topological invariants in the adiabatic approximation change into geo\-metric quantities in the exact factorization.  
Since geometric phases can take any value between $0$ and $2\pi$, not only $0$ or $\pi$, the molecular geometric phase of the current-carrying state $|\Psi_+\rangle$ decreases \textit{continuously} from $\pi$ to a value slightly less than $\pi$ as $M$ is reduced from infinity to some large value.  The molecular geometric phase of a current-carrying state remains finite even though the exact potential energy surface is everywhere smooth, proving that nonanalyticity is not a necessary condition for nonzero molecular geometric phase as previously believed.  Whether there might be other topological contributions to the molecular geometric phase, e.g.~from nodes of $\chi(R)$, is an open question. 

Unlike the Longuet-Higgins phase, which is only path dependent insofar as it depends on the winding number of the path around the conical intersection, the exact molecular geometric phase (like the Berry phase) is truly path dependent.   The integral $\f{1}{\hbar}\oint_{\mathcal{C}} \mathbf{A}_{\mu}\cdot d\R_{\mu}$ is the expression for the flux of a smeared-out flux tube through a surface bounded by the path $\mathcal{C}$.  Since the surface does not enclose all of the flux of the smeared-out flux tube, the integral is not quantized to $0$ or $\pi$.  In the limit $M\rightarrow\infty$, the smeared-out flux tube shrinks to a line and the exact geometric phase approaches the adiabatic value $0$ or $\pi$.  Degeneracy is the crucial factor allowing us to construct current-carrying states with nontrivial Berry curvature.  This is the essential difference with respect to the model studied in Ref.~\onlinecite{min2014}, which does not have a degeneracy following from pseudorotational symmetry because two of the three nuclei are held fixed.

The model studied here provides instructive examples of nontrivial molecular geometric phase and induced vector potentials and exact 2D potential energy surfaces.  Choosing a real-valued state from the degenerate ground state manifold leads to an exact potential energy surface that necessarily breaks the threefold symmetry of the model Hamiltonian.  This is in contrast to the adiabatic case, where there is a single symmetric potential energy surface for all states belonging to a degenerate manifold.  On the other hand, threefold symmetry can be preserved by choosing a complex current-carrying ground state, but then the nuclear Schr\"odinger equation must contain a nontrivial vector potential.  Thus, it appears that choosing a current-carrying state leads to a potential energy surface that is closer to the familiar smooth and symmetric adiabatic potential energy surfaces.

Nontrivial induced vector potentials can occur in any molecular system due to the degeneracy associated with rotational symmetry, although in some cases they may be negligible.  It will be necessary to account for induced vector potentials in some time dependent problems, e.g.~excitations to degenerate excited states with significant nuclear currents.  Methods to exploit the exact factorization in coupled electron-nuclear dynamics are under active development,\cite{agostini2015} but so far have only been applied to one-dimensional systems, where $\mathbf{A}_{\mu}$ is trivial. The physical role of the induced vector potential is to provide a contribution $\mathbf{A}_{\mu} |\chi|^2$ to the nuclear current which the gradient term $\hbar\: \mathrm{Im}\chi^* \nabla_{\mu} \chi$ is not able to provide.  

\textit{Note added in manuscript} -- Englman has applied the exact factorization to the model of Longuet-Higgins and coworkers\cite{longuet-higgins1958} in a paper\cite{englman2015} that appeared after our manuscript was submitted.  The model is a special case of the one studied here, but Ref.~\onlinecite{englman2015} considered only the adiabatic (Longuet-Higgins) phase, 0 or $\pi$, and not the exact geometric phase defined by Eqs.~(\ref{eq:geometricphase}) and (\ref{eq:A}).

\begin{acknowledgments}
We acknowledge helpful discussions with Seung Kyu Min.
\end{acknowledgments}

\appendix

\section{Hamiltonian matrix elements\label{app:A}}

All radial integrals can be evaluated analytically with the recursion relations for the Laguerre polynomials.\cite{koizumi1994}  We start with the first term $\hat{T}^{\rm rad}$ of Eq.~(\ref{eq:kinetic}), which obeys the selection rule $\Delta m \equiv m_2-m_1=0$. We distinguish three cases: (i) $\Delta N \equiv N_2-N_1 = 0$, (ii) $\Delta N = \pm 1$ and (iii) $|\Delta N|>1$, where $N_i=(n_i-|m|)/2$.  
In case (i) we find
\begin{align*}
\langle n m | \hat{T}^{\rm rad} | n m \rangle = \hbar \Omega \Big(N+\f{1}{2}\Big) {;}
\end{align*}
in case (ii),
\begin{align*}
\langle n_1 m | \hat{T}^{\rm rad} | n_2 m \rangle = \f{\hbar \Omega}{2} \f{(\mathrm{max}(N_1,N_2))^{3/2}}{\sqrt{|m|+\mathrm{max}(N_1,N_2)}}   {;}
\end{align*}
and in case (iii), 
\begin{align*}
\langle n_1 m | \hat{T}^{\rm rad} | n_2 m \rangle &=  
-\f{\hbar \Omega}{2} |m| \sqrt{\f{\mathrm{max}(N_1,N_2)!}{\mathrm{min}(N_1,N_2)!}}  \nn \\
&\times \sqrt{\f{(|m|+\mathrm{min}(N_1,N_2))!}{(|m|+\mathrm{max}(N_1,N_2))!}} {.} 
\end{align*}
We have defined the frequency $\Omega=\sqrt{\mathcal{K}/\mathcal{M}}$.
The matrix elements of the second term of Eq.~(\ref{eq:kinetic}), $\hat{T}^{\rm ang}$, obey the selection rule $\Delta m=0$.  They are the same as those for case (iii) of $\hat{T}^{\rm rad}$, but with opposite sign, 
\begin{align*}
\langle n_1 m | \hat{T}^{\rm ang} | n_2 m \rangle &= -\langle n_1 m | \hat{T}^{\rm rad} | n_2 m \rangle \quad \textrm{[case (iii)]} {.}
\end{align*}

The matrix elements of the internuclear repulsion obey the selection rules $\Delta m = 0$ and $\Delta N = 0,\pm 1$.  If $\Delta N=0$, we have
\begin{align*}
\langle n m | \hat{V}_{nn} | n m \rangle = \f{\hbar \Omega}{2} (n+1) .
\end{align*}
and if $\Delta N=\pm 1$, we have
\begin{align*}
\langle n_1 m | \hat{V}_{nn} | n_2 m \rangle = -\f{\hbar \Omega}{2}\!\sqrt{\mathrm{max}(N_1,N_2) (|m| + \mathrm{max}(N_1,N_2))}  {.}
\end{align*}

The electron-nuclear coupling $\hat{H}_{en}$ is the only term which couples states with different values of $m$.  To simplify the evaluation of its matrix elements, we first change to the nuclear coordinate representation
\begin{align}
\langle a_1 n_1 m_1 | \hat{H}_{en} | a_2 n_2 m_2 \rangle 
= \int_0^{2\pi} \f{d\eta}{2\pi} \int_0^{\infty} Q dQ \rho_{n_1 m_1}(Q)  \nn \\
\times H^{en}_{a_1 a_2}(Q\eta) \rho_{n_2 m_2}(Q) e^{i(m_2-m_1) \eta}  {.} \label{eq:Hen:2}
\end{align}
Using Eqs.~(\ref{eq:Hen:1}) and (\ref{eq:linearization}), $H^{en}_{a_1 a_2}(Q\eta)=\langle a_1 Q \eta | \hat{H}_{en} | a_2 Q \eta \rangle$ is divided into the following three contributions:
\begin{align}
% H^{en}_{a_1 a_2}(Q\eta) &= I_{a_1 a_2} + J_{a_1 a_2} \: g\, Q\cos\eta + K_{a_1 a_2}\: g\, Q \sin\eta {.} 
H^{en}(Q\eta) &= I + J \: g\, Q\cos\eta + K\: g\, Q \sin\eta {.} 
\label{eq:Hen:3}
\end{align}
$I$, $J$ and $K$ are $8\times 8$ matrices reported in Appendix~\ref{app:B}.  The matrix elements of $Q \cos\eta$ and $Q \sin\eta$ can be recovered from the following matrix elements of $Q e^{i\eta}$ and $Q e^{-i\eta}$, which were derived in Ref.~\onlinecite{longuet-higgins1958}:
\begin{align*}
\langle n,m | Qe^{-i\eta} |n+1,m+1\rangle &= \langle n+1,m+1|Qe^{i\eta}|n,m\rangle \\ &= \sqrt{\f{n+m+2}{2}} \\
\langle n,m | Q e^{-i\eta} |n-1,m+1\rangle &= \langle n-1,m+1|Qe^{i\eta}|n,m\rangle \\ &= \sqrt{\f{n-m}{2}} {.}
\end{align*}

\section{Electron-nuclear coupling matrices \label{app:B}}

As a function of $(Q,\eta)$, the electron-nuclear coupling can be represented as shown in Eq.~(\ref{eq:Hen:3}).  To see how parity symmetry affects the matrices $I$, $J$ and $K$, it is convenient to express them in the following real-valued electronic basis: 
\begin{align}
|a'\rangle &= \f{1}{\sqrt{2}i} (|a\rangle - |b\rangle)  \nn \\
|b'\rangle &= \f{1}{\sqrt{2}} (|a\rangle + |b\rangle)  \nn \\
|c'\rangle &= \f{1}{\sqrt{2}i} (|c\rangle - |d\rangle)  \nn \\
|d'\rangle &= \f{1}{\sqrt{2}} (|c\rangle + |d\rangle)  \nn \\
|e'\rangle &= \f{1}{\sqrt{2}i} (|e\rangle - |f\rangle)  \nn \\
|f'\rangle &= \f{1}{\sqrt{2}} (|e\rangle + |f\rangle)  \nn \\
|g'\rangle &= -i|g\rangle \nn \\
|h'\rangle &= -|h\rangle {.}
\end{align}
The states $|a'\rangle$, $|c'\rangle$, $|e'\rangle$ and $|g'\rangle$ are odd with respect to the reflection $Q_3\rightarrow -Q_3$, while $|b'\rangle$, $|d'\rangle$, $|f'\rangle$ and $|h'\rangle$ are even.
In the $|\alpha'\rangle$ basis, we have
\begin{align*}
I &= \mathrm{diag}(-3,-3,+3,+3,0,0,0,0) {,}
\end{align*}
\begin{align*}
J &= \LB \bar{cccccccc}
1 & 0 & 0 & 0 & +\f{1}{2} & 0 & -\f{\sqrt{3}}{2} & 0 \\
0 & -1 & 0 & 0 & 0 & -\f{1}{2} & 0 & -\f{1}{2} \\
0 & 0 & -1 & 0 & -\f{1}{2} & 0 & +\f{\sqrt{3}}{2} & 0 \\ 
0 & 0 & 0 & 1 & 0 & +\f{1}{2} & 0 & -\f{1}{2} \\
+\f{1}{2} & 0 & -\f{1}{2} & 0 & 0 & 0 & 0 & 0 \\
0 & -\f{1}{2} & 0 & +\f{1}{2} & 0 & 0 & 0 & -2 \\ 
-\f{\sqrt{3}}{2} & 0 & +\f{\sqrt{3}}{2} & 0 & 0 & 0 & 0 & 0 \\
0 & -\f{1}{2} & 0 & -\f{1}{2} & 0 & -2 & 0 & 0
\ear \RB {,}
\end{align*}
and
\begin{align*}
K &= \LB \bar{cccccccc}
0 & -1 & 0 & 0 & 0 & -\f{1}{2} & 0 & +\f{1}{2} \\
-1 & 0 & 0 & 0 & -\f{1}{2} & 0 & -\f{\sqrt{3}}{2} & 0 \\
0 & 0 & 0 & +1 & 0 & +\f{1}{2} & 0 & +\f{1}{2} \\ 
0 & 0 & +1  & 0 & +\f{1}{2} & 0 & +\f{\sqrt{3}}{2} & 0 \\
0 & -\f{1}{2} & 0 & +\f{1}{2} & 0 & 0 & 0 & 2 \\
-\f{1}{2} & 0 & +\f{1}{2} & 0 & 0 & 0 & 0 & 0 \\ 
0 & -\f{\sqrt{3}}{2} & 0 & +\f{\sqrt{3}}{2} & 0 & 0 & 0 & 0 \\
+\f{1}{2} & 0 & +\f{1}{2} & 0 & 2 & 0 & 0 & 0
\ear \RB {.}
\end{align*}
The matrix $J$ couples states with the same parity because $\cos\eta$ is even under reflection.  The matrix $K$ couples states of different parity because $\sin\eta$ is odd.

\bibliography{bibliography}

\end{document}